\documentclass[aps,twocolumn,nofootinbib]{revtex4}
\usepackage{amsfonts}
\usepackage{amsmath}
\usepackage{amssymb}
\usepackage{graphicx}
\newtheorem{theorem}{Theorem}

\newtheorem{definition}{Definition}

\newtheorem{lemma}{Lemma}

\newenvironment{proof}[1][Proof]{\noindent\textbf{#1.} }{\ \rule{0.5em}{0.5em}}

\begin{document}

\title{Ontological models and the interpretation of contextuality}

\author{Nicholas Harrigan}
\affiliation{QOLS, Blackett Laboratory, Imperial College London,
Prince Consort Road, London SW7 2BW, United Kingdom}

\author{Terry Rudolph}
\affiliation{QOLS, Blackett  Laboratory, Imperial College London,
Prince Consort Road, London SW7 2BW, UK} \affiliation{Institute for
Mathematical Sciences, Imperial College London, \\53 Exhibition
Road, London SW7 2BW, UK}

\begin{abstract}
Studying the extent to which realism is compatible with quantum
mechanics teaches us something about the quantum mechanical
universe, regardless of the validity of such realistic assumptions.
It has also recently been appreciated that these kinds of studies
are fruitful for questions relating to quantum information and
computation. Motivated by this, we extend the \textit{ontological
model formalism} for realistic theories to describe a set of
theories emphasizing the role of measurement and preparation devices
by introducing `hidden variables' to describe them. We illustrate
both the ontological model formalism and our generalization of it
through a series of example models taken from the literature. Our
extension of the formalism allows us to quantitatively analyze the
meaning contextuality (a constraint on successful realistic
theories), finding that - taken at face-value - it can be realized
as a natural interaction between the configurations of a system and
measurement device. However, we also describe a property that we
call deficiency, which follows from contextuality, but does not
admit such a natural interpretation. Loosely speaking, deficiency
breaks a symmetry between preparations and measurements in quantum
mechanics. It is the property that the set of ontic states which a
system prepared in quantum state $|\psi\rangle$ may actually be in,
is strictly smaller than the set of ontic states which would reveal
the measurement outcome $|\psi\rangle\langle\psi|$ with certainty.
\end{abstract}
\maketitle

\section{Introduction \label{SEC:introduction}}

Quantum mechanics is famously plagued by certain conceptual
problems, the resolution of which drive attempts to understand the
theory. These attempts have resulted in the appearance of a diverse
number of interpretations of quantum mechanics - ideas about how to
relate mathematical objects from the theory to some picture of (or
viewpoint regarding) physical reality. Somewhat incredibly, there is
still not even a consensus  on precisely which features of quantum
mechanics are the source of these conceptual problems.

One approach that has been advocated is to simply deny the need for
understanding quantum mechanics in terms of a metaphysical picture
of reality at all. We will have nothing to say about such a
dismissive approach in this paper. However, if, as will be assumed
here, it is desirable to understand quantum mechanics in a realistic
framework, then many possibilities arise. The simplest realistic
approach is to simply assert that the quantum state itself is in
one-to-one correspondence with reality. This, as Einstein and others
have emphasized \cite{harrigan_spekkens,Howard_einst_short}, entails
accepting a view of physical reality with arguably quite undesirable
features (e.g. violent nonlocality, discontinuous dynamics,
ambiguous emergence of a classical ontology etc.).

Our goal in this paper is to lay out and expand upon a framework and
a language in which (almost) any theory attempting to correlate
quantum mechanics to a picture of reality can be formulated. This
framework, first introduced in \cite{spekkens_con}, includes the
just-mentioned possibility that the quantum state \textit{is} the
state of reality. However, as emphasized in
\cite{harrigan_spekkens}, it also includes possibilities wherein the
quantum state is supplemented by some ``hidden variables''.
Regardless of whether there are such hidden variables besides
quantum states, it is possible that one might be able to interpret
the quantum state \textit{epistemically}
\cite{Fuchs,FuchsJmodopt,Leiferarxiv,Leiferpra} - that is, in terms
of probability distributions over some space (see
\cite{toy_theory,tr_model} for explicit examples of such an
epistemic construction). If a theory for the reality underpinning
quantum mechanics can be formulated in the general terms we propose
then we refer to it as an \textit{ontological model}. Following
\cite{spekkens_con}, the ``true states of reality'' posited by the
model will be called ``ontic states''. The terminology is chosen to
emphasize that while such theories are not necessarily
`\textit{hidden} variable theories', they do attempt to formulate a
picture of physical reality consistent with quantum mechanics.

Although one might not expect an ontological model to precisely
follow the laws of classical mechanics, there are certain features,
commonplace in classical physics, that one would hope could be
retained - for instance, conservation laws and locality. Amazingly,
Bell's theorem \cite{Bell_locality} shows that locality must be
abandoned in any theory whatsoever that describes our universe
\cite{norsen}, including, of course, any ontological model. This
feat of generality rested on Bell's ability to abstract generic
features possessed by all realistic models. Consolidating and
extending such generality is one goal of the ontological model
formalism that we build upon. Besides nonlocality, the other primary
non-classical feature which any attempt at explaining quantum
mechanics in a realistic framework must contend with is
\emph{contextuality}. Contextuality, first considered for quantum
mechanics by Kochen and Specker \cite{Ks} and then extended to deal
with arbitrary theories by Spekkens \cite{spekkens_con}, is much
less understood and appreciated than nonlocality. Increasing the
generality of the ontological model formalism also works towards a
second goal of this paper; to elucidate the precise manner in which
contextuality must manifest itself in all such models.

In addition to the above motivations, which originate from
foundational considerations, a second series of motivations for this
research stem from practical issues in the field of quantum
information theory. Precise formulations of a spectrum of realistic
theories potentially underpinning quantum mechanics are of use to
work in this field, regardless of their metaphysical consequences.
Such formulations allow us to probe and elucidate those features of
quantum mechanics distinguishing it from classical realistic
theories - the theories upon which all of classical information
theory is predicated. While the role of quantum nonlocality (and
entanglement in particular) in distinguishing quantum and classical
information theory has been much speculated upon, contextuality has
received far less attention in this regard \cite{spekkens_con}. We
believe this neglect to be a serious mistake. Furthermore, Aaronson
\cite{scott} has recently discussed how one can define complexity
classes in terms of the increased computational power one might
expect if one were able to access individual ontic states (and
obtain more information about a system than the quantum formalism
itself allows). We show in Sec.~\ref{SEC:example_Aaronsonmodel} how
the theories considered by Aaronson can be expressed in the
ontological model formalism. In particular, it then becomes clear
that not all ontological models yield the computational advantages
that Aaronson identifies. The paper begins in
Sec.~\ref{SEC:om_intro} by presenting the ontological model
formalism as it can be applied to quantum systems. In the next
section a variety of ontological models, chosen to illustrate the
breadth of possibilities, are discussed. These models include the
two famous examples from Bell's papers
\cite{Bell_context,Bell_locality}, Kochen and Specker's
non-contextual model of a qubit \cite{Ks}, Aaronson's model
\cite{scott}, a model (due to Beltrametti and Bugajski) which takes
the quantum state itself as real \cite{beltrametti_bugajski}, and
some interesting models of Aerts \cite{Aerts85,Aerts91}. It will
quickly become clear that the formalism of Sec.~\ref{SEC:om_intro}
needs some augmentation, particularly if we want to be able to
discuss the physical reality of preparation and measurement devices
themselves (as any posited realistic theory of the whole universe
should). In Sec.~\ref{SEC:ont_meas_prep} we therefore undertake
formulating such an extension, and find that several interesting new
possible features arise which can distinguish different ontological
models. We then turn to a deeper examination of how ontological
models deal with the Kochen-Specker theorem. In doing so we identify
a property we term \textit{deficiency}, which all ontological models
possess, and which forms the subject of
Sec.~\ref{SEC:deficiency_intro}. Deficiency involves the explicit
breaking of the symmetry between preparations and measurements that
is enjoyed by quantum mechanics (e.g. that a preparation of a
quantum state can be achieved by a projective measurement onto that
state of a suitable system.) We also show how deficiency elucidates
the fact that measurements in an ontological model must be
disturbing.



\section{Ontological models and quantum mechanics \label{SEC:om_intro}}

The formalism of quantum mechanics is well known and relatively
unambiguous, but opinions are varied on just what this formalism is
meant to describe, i.e. how it corresponds in some sense to reality.
One of the most popular views is the operational one \cite{peres};
wherein the only concern of the theory is to reproduce outcomes of
various experimental procedures employed by a scientist. The
question of how quantum mechanics relates to reality is then taken
to be outside the theories scope.

\begin{figure}[t]
\includegraphics[scale=0.4]{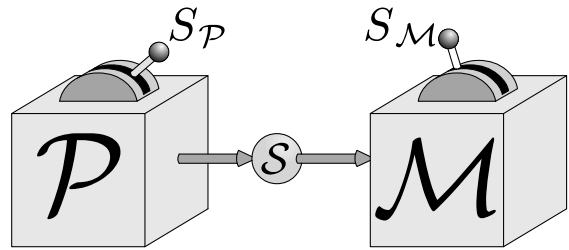}\caption{The `$\mathcal{P}\rightarrow\mathcal{M}$' paradigm for operational theories.}%
\label{FIG:prepmeas}%
\end{figure}

This approach to quantum theory is one of two views that are
commonly held (often implicitly). The other frequently maintained
position is that reality is \textit{completely} described by the
quantum state - so that within its domain of validity it is the end
of the story. This is implicitly a realistic view, and therefore one
that will be incorporated in the ontological model formalism. Unless
otherwise stated, when referring to quantum mechanics we will always
have in mind an operational interpretation of the theory. If it is
to be experimentally verified, any such operational theory needs to
be able to describe the paradigm illustrated in
Fig.~\ref{FIG:prepmeas}. A system $\mathcal{S}$, initially interacts
with a preparation device $\mathcal{P}$ which is configured
according to some macroscopically determinable setting
$S_{\mathcal{P}}$. This setting is manipulated in order to alter
what the state of $\mathcal{S}$ will be as it leaves $\mathcal{P}$.
$\mathcal{S}$ then travels towards a measurement
device\footnote{Although it has a natural representation in the
ontological model formalism, we will not need to consider the
possibility of a transformation acting on a system between
$\mathcal{P}$ and $\mathcal{M}$.}, $\mathcal{M}$, configured
according to some setting $S_{\mathcal{M}}$. $\mathcal{M}$ will then
register some particular outcome dependent on both the state of
$\mathcal{S}$ and the setting $S_{\mathcal{M}}$.

We can (for our purposes) define operational quantum mechanics by
how it determines measurement statistics within this kind of
$\mathcal{P}\rightarrow\mathcal{M}$ scenario,

\begin{definition}
Quantum Mechanics is a theory that describes Fig.~\ref{FIG:prepmeas}
by associating a density operator $\rho_{S_{\mathcal{P}}}$ (on a
suitably chosen Hilbert space $\mathcal{H}$) with a preparation
procedure $S_{\mathcal{P}}$, and a positive operator valued measure
(POVM) $\left\{E_k\right\}_k$ with the measurement procedure
$S_{\mathcal{M}}$, there being one `POVM effect', $E_k$ for each of
the possible measurement outcomes. The quantum prediction for the
probability of the $k^{th}$ outcome in $S_{\mathcal{M}}$ occurring
conditioned on a preparation $S_{\mathcal{P}}$ is then given by the
Born rule,
$\text{Pr}\left(k|S_{\mathcal{P}},S_{\mathcal{M}}\right)=\textrm{tr}\left(E_k\rho_{S_{\mathcal{P}}}\right)$.
\end{definition}

Of course, special cases of this formalism are that quantum
mechanics associates rays $|\psi\rangle\in\mathcal{H}$ with pure
state preparations and projection operators with sharp (rank one)
measurements, which can be thought of as `testing' whether or not a
system is in a particular pure state.

Quantum mechanics, defined in this operational way, is exceedingly
successful at reproducing observed statistics, but it doesn't give
us any picture of what ``really'' goes on inside a system when
experimental procedures are performed on it. In Newtonian mechanics
one deals with measurements, preparations and evolutions of a
particle's position, and this position is posited to be
ever-existing, simply revealed to us by measurement, so the theory
is quite clear on how its predictions relate to reality. In
comparison, quantum mechanics deals with transformations of state
vectors and no prior relation is specified between these state
vectors and reality.

A realistic view of quantum mechanics adds to this picture with the
aim of providing a link between the quantum mechanical formalism and
an underlying reality. There is of course no unique way in which one
might achieve this kind of realistic interpretation, and in fact
many such constructions have been given to date, the most famous
surely being Bohmian mechanics \cite{bohm,bohmsurvey}. In Bohmian
mechanics the quantum state of a particle and a specification of its
position are taken to correspond directly to elements of the
`underlying reality'. Other attempts at realistic constructions can
be found in
\cite{scott,Bell_locality,toy_theory,beltrametti_bugajski,Ks,griffiths_book,spon_collapse}
We provide a more detailed consideration of a representative
selection of these constructions (and show how they can be expressed
in the formalism we use) in Sec.~\ref{SEC:example_models}.

To identify features common to these realistic constructions we
would like a general language which allows us to abstract away the
specific details of any one particular realistic view. We use the
term ontological model to refer to a very natural, although
non-exhaustive, formalism which does just this job. For the
remainder of the paper we will implicitly restrict our attention to
those realistic constructions expressible in this
formalism\footnote{One of the reasons that the ontological model
formalism does not exhaust \textit{all} of the possible realistic
ways of interpreting quantum mechanics is because it employs several
assumptions about the behavior of reality. This will become apparent
when we consider extending the conventional formalism in
Sec.~\ref{SEC:contextuality_interp}.}, referring to them as
\textit{ontological models} \cite{spekkens_con}. So what will a
general ontological model look like? Any such model should pick up
precisely where operational quantum mechanics leaves off, and
specify just what it is that a quantum state allows us to infer
about the real state of a system. The model can then be filled out
by considering how each of the operations in Fig.~\ref{FIG:prepmeas}
are taken to act on these hypothesized \textit{real states} of the
system. We would expect that acting a preparation procedure
$\mathcal{P}$ on a system $\mathcal{S}$ would configure
$\mathcal{S}$ so that it possesses some particular real state after
the preparation. A measurement procedure $\mathcal{M}$ would then
correspond to some kind of interaction with $\mathcal{S}$ - an
interaction tailored to be such that $\mathcal{M}$ registers one or
another measurement outcome dependent on the prior real state of
$\mathcal{S}$.

An ontological model quantifies these realistic notions, by
introducing a set $\Lambda$ of \textit{ontic states} $\lambda$ to be
associated with $\mathcal{S}$. These constitute a complete
description of whatever reality the model takes to underpin the
system, so that a specification of $\lambda$ is a complete
description of any attributes that $\mathcal{S}$ might possess. The
precise form taken by $\Lambda$ will depend on the particular
ontological model under consideration and the nature of the
underlying reality that it introduces. In the simplest possible
realistic interpretation, we can take quantum states to be direct
and complete descriptions of reality. Then we obtain an ontological
model in which the ontic state space $\Lambda$ is precisely equal to
the projective Hilbert space of $\mathcal{S}$, i.e.
$\Lambda=\mathcal{PH}$. More generally however it might be the case
that $\Lambda\neq\mathcal{PH}$. Then either the quantum state is not
a complete description of reality and must be supplemented by extra
`hidden' variables ($\mathcal{PH}\subset\Lambda$), or the quantum
state does not play a realistic role at all
($\mathcal{PH}\nsubseteq\Lambda$), and must simply represent our
\textit{state of knowledge} of the real state of $\mathcal{S}$. For
example, in Bohmian Mechanics, elements of $\Lambda$ consist of a
specification of the system's quantum state \textit{and} a
specification of the system's position and therefore $\Lambda$ takes
the form of a cartesian product
$\Lambda=\mathcal{PH}\times\mathbb{R}^3$.

So the state space $\Lambda$ provides a description of the real
state of the \textit{system}, $\mathcal{S}$. Preparation and
measurement devices $\mathcal{P}$ and $\mathcal{M}$, ultimately
being physical systems, should also be describable in terms of their
own set of ontic states. However, the ontological model formalism
has traditionally been restricted to a realistic description of
$\mathcal{S}$ alone, simplifying matters by treating $\mathcal{P}$
and $\mathcal{M}$ as external to the theory. In
Sec.~\ref{SEC:contextuality_interp} we show how to extend the
ontological model formalism to also provide ontological treatments
for these devices, allowing us to consider a wider class of models
and affording an insight into the manifestation of contextuality in
realistic theories. For now, however, we will restrict our attention
to the traditional formulation of providing a realistic description
of the \textit{system} only.

In this simplified picture (wherein we neglect ontological
descriptions of $\mathcal{P}$ and $\mathcal{M}$) how does an
ontological model quantify preparations and measurements in terms of
operations on the real states of the \textit{system}, $\mathcal{S}$?
In general, performing a preparation with setting $S_{\mathcal{P}}$
will result in the system $\mathcal{S}$ being prepared in some
particular ontic state $\lambda\in\Lambda$. Simply knowing
$S_{\mathcal{P}}$ may, however, be insufficient information to
deduce precisely which $\lambda$ a system is in. Thus, in general,
an ontological model will associate a probability distribution
$\mu\left(\lambda|S_{\mathcal{P}}\right)$ over $\Lambda$ with
preparation procedure $S_{\mathcal{P}}$. This distribution encodes
our \textit{epistemological uncertainty} as to the precise
ontological configuration of $\mathcal{S}$, and so we refer to it as
an \textit{epistemic state}. Note that since a system must be
described by \textit{some} $\lambda\in\Lambda$ we will require that,

\begin{equation}
\int_{\Lambda}{d}\lambda\:\mu(\lambda|S_{\mathcal{P}})=1.
\label{mu_norm}
\end{equation}

Associating $|\psi\rangle$ with a probability distribution is
obviously compatible with the notion of quantum states having no
direct relation to the ontic states, but it is also consistent with
quantum states being taken to be precisely the ontic states
themselves. To allow for this we need only take
$\Lambda=\mathcal{PH}$ and write
$\mu(\lambda|\psi)=\delta(\lambda-\lambda_{\psi})$ with $\delta$
being the Dirac delta function and $\lambda_{\psi}$ the unique ontic
state associated with preparation settings consistent with
$|\psi\rangle$. Hence the view where quantum states are taken to be
complete descriptions of reality can easily be expressed in the
ontological model formalism. In the next section we will see an
explicit example of a model that achieves this.

Consider now a measurement wherein $\mathcal{M}$ is configured
according to some setting $S_{\mathcal{M}}$. The outcome of this
measurement will be determined by the ontic state $\lambda$ of the
system and how it interacts with $\mathcal{M}$ (a point which we
elaborate on in Sec.~\ref{SEC:contextuality_interp}). Now the most
general possibility is that $\lambda$ might only
\textit{probabilistically} determine a measurement outcome.
Following \cite{spekkens_con}, we refer to models wherein even a
complete description of reality only allows one to make
probabilistic predictions, as being \textit{outcome
indeterministic}. Conversely if the ontic state $\lambda$ of
$\mathcal{S}$ is sufficient to completely determine a measurement's
outcome then we call the model \textit{outcome deterministic}. To
allow for both these possibilities we therefore represent the
$k^{th}$ outcome of a measurement performed according to
$S_{\mathcal{M}}$ by a distribution
$\xi\left(k|\lambda,S_{\mathcal{M}}\right)$ over $\Lambda$, telling
us the probability that a given $\lambda\in\Lambda$ will yield the
$k^{th}$ outcome. We refer to such distributions as
\textit{indicator functions} (considered as functions of $\lambda$).
In outcome deterministic models,
$\xi\left(k|\lambda,S_{\mathcal{M}}\right)\in\left\{0,1\right\}$ -
so that the indicator functions are idempotent, i.e. we have
$\xi^2(k|\lambda,S_{\mathcal{M}})=\xi(k|\lambda,S_{\mathcal{M}})$
for all $\lambda$. Where might the probabilities appearing in
outcome indeterministic models arise from? There are two
possibilities. Firstly they could occur because of our failure to
take into account the precise ontological configurations of either
$\mathcal{P}$ or $\mathcal{M}$, a possibility which we address in
Sec.~\ref{SEC:contextuality_interp}. Alternatively it could be that
the probabilities are an inherent property of the reality described
by the model, so that even if one had complete knowledge of the
configuration of the whole universe, one would be unable to make any
certain statements about the system's future configuration.

Since one or the other outcome of any measurement $S_{\mathcal{M}}$
must occur - no matter what $\lambda$ describes $\mathcal{S}$ - we
have,
\begin{equation}
\sum_k\xi(k|\lambda,S_{\mathcal{M}})=1\:\:\:\forall\lambda.
\label{ind_fns_sum_one}
\end{equation}
The settings $S_{\mathcal{P}}$ and $S_{\mathcal{M}}$ will play a
crucial role in many of our discussions. Clearly different settings
$S_{\mathcal{P}}$ can describe situations in which $\mathcal{P}$ is
set - within an operational quantum mechanical description - to
prepare a system according to different density operators.
Similarly, different settings $S_{\mathcal{M}}$ can describe cases
where $\mathcal{M}$ is set to implement different POVM measurements.
However, there are also many distinct settings of $\mathcal{P}$ and
$\mathcal{M}$ consistent with a quantum mechanical description given
by the \textit{same} density operator or POVM. The settings will
then specify different instances of some other extraneous property
of $\mathcal{P}$ or $\mathcal{M}$. We will later see that there
exist quantitative extraneous properties which, although not
altering the POVM implemented, must alter the indicator function
used by an ontological model. Thus the quantum mechanical POVM
description of a measurement can actually be thought of as being a
function $E(S_{\mathcal{M}})$ of the measurement setting of
$\mathcal{M}$ - in that each POVM corresponds in general to a
certain \textit{set} of settings of $\mathcal{M}$. Hence although
specifying $S_{\mathcal{M}}$ will uniquely fix a POVM $E$, knowledge
of only $E$ may be insufficient to completely determine
$S_{\mathcal{M}}$. The full setting, $S_{\mathcal{M}}$, of
$\mathcal{M}$ is referred to as the \textit{measurement context} (a
term we define in more detail later). Hence, fully specifying the
measurement context may require stating not just a POVM $E$, but
also some `extra' information which completely determines
$\mathcal{M}$'s setting\footnote{Note also that on occasion we will
lazily refer to a POVM $E$ as defining a measurement setting.
Strictly speaking of course, we mean to say ``a measurement setting
that is described \textit{in quantum mechanics} by a POVM $E$''.}.
So although we may occasionally write $\xi(k|\lambda,E)$, we should
really make explicit the precise setting $S_{\mathcal{M}}$ by
writing either $\xi(k|\lambda,S_{\mathcal{M}})$ or (if we still want
to make clear the POVM), $\xi(k|\lambda,E,S_{\mathcal{M}})$.

Similarly, a density operator $\rho$ may be compatible with many
preparation settings $S_{\mathcal{P}}$, and so although we will
often write epistemic states as $\mu(\lambda|\rho)$, we should
really express them in the form $\mu(\lambda|S_{\mathcal{M}})$ or
$\mu(\lambda|\rho,S_{\mathcal{M}})$.

To summarize then, for the purposes of this paper, we can define an
ontological model by the following criteria,

\begin{definition}
An ontological model posits an ontic state space $\Lambda.$ The
probability of the ontic state being $\lambda$, given the
preparation procedure $S_{\mathcal{P}}$ is denoted by a probability
distribution which we refer to as an \textit{epistemic state},
$\mu(\lambda|S_{\mathcal{P}})$. The probability of measurement
outcome $k$ occurring given that the ontic state is $\lambda$ and
the measurement procedure was $S_{\mathcal{M}}$ is given by an
indicator function, written $\xi(k|\lambda,S_{\mathcal{M}})$ (with
$\xi^2(k|\lambda,S_{\mathcal{M}})=\xi(k|\lambda,S_{\mathcal{M}})$ in
outcome deterministic models). We then demand that a successful
ontological model of quantum mechanics should reproduce the required
statistics by satisfying,
\begin{equation}
\int\mathrm{d}\lambda\:\xi(k|\lambda,E,S_{\mathcal{M}})\mu(\lambda|\rho,S_{\mathcal{P}})=\mathrm{tr}\left(\rho{E}_k\right).
\label{ont_mod_qm_stats}
\end{equation}
\end{definition}

Seen from the viewpoint of an ontological model, a quantum
mechanical picture of reality generally corresponds to a
coarse-graining over the ontic states. By explicit construction, all
ontological models will yield the same statistical predictions at
this coarse-grained `quantum level'. In many models, complete
knowledge of the ontic configuration of a system would lead one to
make predictions differing in some way from those of quantum theory.
Serious advocates of ontological models might claim that the reason
we do not see these deviations from quantum predictions is because
our current experiments are still too `coarse-grained' to be able to
operate on the level of individual ontic states. Another
possibility, is that ontological models might inherently exhibit a
restriction such that maximal possible knowledge of a system's
ontological configuration is always \textit{incomplete} knowledge
\cite{Fuchs}. The ontic states describing a system would then, to
some extent, be `inherently unknowable'. Although such a
restriction-of-knowledge principle has been shown to have the
potential to reproduce many characteristic features of quantum
mechanics \cite{toy_theory}, it is not a \textit{necessary} feature
of all ontological models.

Even though manipulation of individual ontic states is potentially
forbidden (either technologically or inherently), we will still have
occasion to consider the predictions that a model would be able to
make if we hypothetically were somehow able to prepare and
distinguish between individual ontic states. In particular we will
find it useful to refer to a kind of equivalence between models,
which we define as follows,

\begin{definition}
An ontological model $\mathcal{O}$ is said to be
\textbf{ontologically equivalent} to a second model
$\tilde{\mathcal{O}}$ if all statistics predicted by
$\tilde{\mathcal{O}}$ are exactly reproduced by model $\mathcal{O}$,
even in cases where one is able to perform preparation and
measurement procedures that distinguish between individual ontic
states. \label{DEF:ont_equiv}
\end{definition}

\begin{figure}[t]
\includegraphics[scale=0.6]{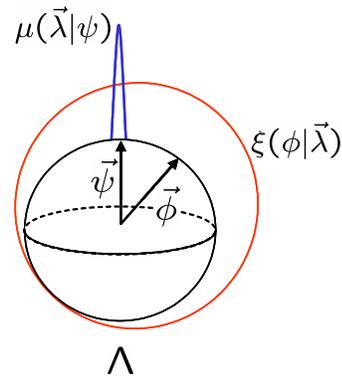}\caption{Illustration of the epistemic
states and indicator functions of the Beltrametti-Bugajski model.}
\label{FIG:bbmodel}
\end{figure}

\section{Examples of ontological models \label{SEC:example_models}}

The formalism that we have described so far is sufficient to
describe many existing ontological models. However, there exist
models which lie outside of its scope because of the way that they
treat the measurement apparatus. In this section we present some
examples of ontological models that show both the utility and
limitations of the standard ontological model formalism\footnote{The
formulations presented here for the Beltrametti-Bugajski model and
the Kochen Specker model first appeared in \cite{harrigan_spekkens}.
Note that the model referred to in \cite{harrigan_spekkens} as
``Bell's model'' is an adaptation (by Mermin \cite{mermin_bell}) of
what we call Bell's \textit{second} model.}. The limitations we
encounter will act as motivations to generalize the formalism, a
task we undertake in Sec.~\ref{SEC:ont_meas_prep}.


\subsubsection{The Beltrametti-Bugajski model \label{SEC:example_BBmodel}}

The model of Beltrametti and Bugajski \cite{beltrametti_bugajski} is
essentially a thorough rendering of what most would refer to as an
orthodox interpretation of quantum mechanics\footnote{Note, however,
that there are several versions of orthodoxy that differ in their
manner of treating measurements. The Beltrametti-Bugajski model is
distinguished by the fact that it fits within the framework for
ontological models we have outlined.}. The ontic state space
postulated by the model is precisely the projective Hilbert space,
$\Lambda=\mathcal{PH}$, so that a system prepared in a quantum state
$\psi$ is associated with a sharp probability
distribution\footnote{Preparations which correspond to mixed quantum
states can be constructed as a convex sum of such sharp
distributions.} over $\Lambda$,
\begin{equation} \mu\left(
\lambda|\psi\right)d\lambda  =\delta\left(
\lambda-\lambda_{\psi}\right)d\lambda,
\end{equation}
where we are using $\psi$ interchangeably to label the Hilbert space
vector and to denote the ray spanned by this vector.
$\lambda_{\psi}$ denotes the unique ontic state associated with the
quantum state $\psi$. Thus the model posits that the different
possible states of reality are simply the different possible quantum
states.

Quantum statistics are reproduced by assuming that the probability
of obtaining an outcome $k$ of a measurement procedure
$S_{\mathcal{M}}$ depends indeterministically on the system's ontic
state $\lambda$ as,
\begin{equation}
\xi\left(  k|\lambda,E,S_{\mathcal{M}}\right)  =\text{tr}\left(
|\lambda\rangle\langle
\lambda|E_{k}\right),\label{BB1}%
\end{equation}
where $\left\vert \lambda\right\rangle \in\mathcal{H}$ denotes the
quantum state associated with $\lambda\in\Lambda$, and where
$E=\left\{ E_{k}\right\}_k$ is the POVM that quantum mechanics
associates with $S_{\mathcal{M}}$. It follows that,
\begin{align}
\mathrm{Pr}\left(  k|E,\psi\right)   &  =\int_{\Lambda}{d}\lambda\text{ }%
{\xi}\left(  k|\lambda,E,S_{\mathcal{M}}\right)  \text{ }\mu(\lambda|\psi)\nonumber\\
&  =\int_{\Lambda}{d}\lambda\text{ tr}\left(  |\lambda\rangle\langle
\lambda|E_{k}\right)  \text{ }\delta\left(
\lambda-\lambda_{\psi}\right)
\label{BB2}\\
&  =\text{tr}\left(  |\psi\rangle\langle\psi|E_{k}\right),\label{BB3}%
\end{align}
and so the quantum statistics are trivially reproduced.

If we restrict consideration to a system with a two dimensional
Hilbert space then $\Lambda$ is isomorphic to the Bloch sphere, so
that the ontic states are parameterized by Bloch vectors of unit
length, which we denote by $\vec{\lambda}.$ \ The Bloch vector
associated with the Hilbert space ray $\psi$ is denoted $\vec{\psi}$
and is defined by $\left\vert \psi\right\rangle \left\langle
\psi\right\vert
=\frac{1}{2}\openone+\frac{1}{2}\vec{\psi}\cdot\vec{\sigma}$ where
$\vec{\sigma }=(\sigma_{x},\sigma_{y},\sigma_{z})$ denotes the
vector of Pauli matrices and $\openone$ denotes the identity
operator.

If we furthermore consider $S_{\mathcal{M}}$ to represent a
\textit{projective} measurement, then it is associated with a
projector-valued measure (PVM) $\{\left\vert \phi \right\rangle
\left\langle \phi\right\vert ,\left\vert \phi^{\perp }\right\rangle
\left\langle \phi^{\perp}\right\vert \}$ or equivalently, an
orthonormal basis $\{\left\vert \phi\right\rangle ,\left\vert
\phi^{\perp }\right\rangle \}$. \ Eq.~(\ref{BB1}) then simplifies
to,
\begin{align}
\xi(  \phi|\vec{\lambda})    & =|\left\langle \phi|\lambda
\right\rangle |^{2}\\
& =\frac{1}{2}\left(  1+\vec{\phi}\cdot\vec{\lambda}\right).
\label{BBif}
\end{align}
Where for brevity, we denote the indicator function $\xi(
1|\vec{\lambda},|\phi\rangle\langle\phi|,S_{\mathcal{M}})$
associated with a projector $|\phi\rangle\langle\phi|$ as $\xi(
\phi|\vec{\lambda})$.

The epistemic states and indicator functions for this two
dimensional case of the Beltrametti-Bugajski model are illustrated
schematically in Fig.~\ref{FIG:bbmodel}.

\subsubsection{The Kochen-Specker model \label{SEC:example_ksmodel}}

\begin{figure}[t]
\includegraphics[scale=0.6]{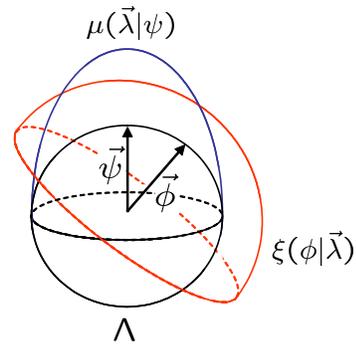}\caption{Illustration of the epistemic
states and indicator functions of the Kochen-Specker model.}
\label{FIG:ksmodel}
\end{figure}

We now consider a model for a two-dimensional Hilbert space due to
Kochen and Specker \cite{Ks}. The ontic state space $\Lambda$ is
taken to be the unit sphere, so that individual ontic states can be
written as unit vectors, $\vec{\lambda}\in\Lambda$. A quantum state
$\psi$ is then associated with the probability distribution,

\begin{equation}
\mu(\vec{\lambda}|\psi)\:\:d\vec{\lambda}=\frac{1}{\pi}\Theta(\vec{\psi}\cdot\vec{\lambda})\:\vec{\psi
}\cdot\vec{\lambda}\:\:d\vec{\lambda}, \label{mu_kochen_specker}
\end{equation}
where $\vec{\psi}$ is the Bloch vector corresponding to the quantum
state $\psi$ and $\Theta$ is the Heaviside step function. This
epistemic state assigns the value $\cos{\theta}$ to all points an
angle $\theta<\frac{\pi}{2}$ from $\vec{\psi}$, and the value zero
to points with $\theta>\frac{\pi}{2}$. This is illustrated in
Fig.~\ref{FIG:ksmodel}.

Upon implementing a measurement procedure $S_{\mathcal{M}}$
associated with a projector $|\phi\rangle\langle\phi|$ a positive
outcome will occur if the ontic state $\vec{\lambda}$ of the system
lies in the hemisphere centered on $\vec{\phi}$, i.e.,
\begin{equation}
\xi(\phi|\vec{\lambda})=\Theta(\vec{\phi}\cdot\vec{\lambda}).\label{xi_ks}
\end{equation}
It can be checked that the overlaps of $\mu(\vec{\lambda}|\psi)$ and
$\xi(\phi|\vec{\lambda})$ then reproduce the required quantum
statistics,
\begin{align}
\int{d}\vec{\lambda}\:\mu(\vec{\lambda}|\psi)\:\xi(\phi|\vec{\lambda})
& =\int{d}\vec{\lambda}\:\frac{1}{\pi}\Theta(\vec{\psi}\cdot
\vec{\lambda})\Theta(\vec{\phi}\cdot\vec{\lambda})\:\vec{\psi}\cdot
\vec{\lambda}\nonumber\\
&  =\frac{1}{2}(1+\vec{\psi}\cdot\vec{\phi})\nonumber\\
&  =\left|  \langle\psi|\phi\rangle\right|  ^{2}.
\end{align}

This model is outcome deterministic, and therefore demonstrates how
one can reproduce quantum statistics solely through a lack of
knowledge about which ontic state a system is prepared in.

\subsubsection{Aaronson's models \label{SEC:example_Aaronsonmodel}}

In a recent paper \cite{scott} Aaronson developed a formalism for
describing a certain class of ontological models in terms of
stochastic matrices. Aaronson then went on to consider the
computational complexity of simulating models from this class.

The idea behind Aaronson's models is to replace the Hilbert space
vector $|\psi\rangle$ describing a quantum system with a vector
$v_{\psi}$ of the amplitudes of the state when written in some
preferred basis $\Omega=\left\{|\omega_i\rangle\right\}_{i=1}^N$,
i.e.
$v_{\psi}=\left[|\langle\psi|\omega_1\rangle|^2,\ldots,|\langle\psi|\omega_N\rangle|^2\right]$.
The action of any unitary transformation on $|\psi\rangle$ is then
mimicked by a map $\mathbb{S}:\Omega\rightarrow\Omega$, represented
by a stochastic matrix $\mathbb{S}$ acting on the vector $v_{\psi}$.
As Aaronson shows in \cite{scott}, such a matrix must depend not
only on the unitary transformation that it attempts to enact, but
also on the particular quantum state that it is to be acted on. Thus
we can write the stochastic matrix intended to reproduce the action
of a unitary $U$ on a state $|\psi\rangle$ as $\mathbb{S}(U,\psi)$.
The specific form of these stochastic matrices is dependent on the
particular hidden variable theory from Aaronson's formalism. In
order to make sure that these theories reproduce quantum mechanical
predictions, the matrices must satisfy,
\begin{equation}
\sum_i\mathbb{S}(U,\psi)_{ji}|\langle\omega_i|\psi\rangle|^2=|\langle\omega_j|U|\psi\rangle|^2.
\label{Aaronson_qmstat}
\end{equation}
In this scheme, any attempt to perform a measurement on
$|\psi\rangle$ in a basis $\mathcal{B}$ other than $\Omega$ is
interpreted as a unitary evolution, $U$, rotating $\psi$ into the
basis $\Omega$ (represented by a relevant stochastic matrix),
followed by a measurement in this preferred basis. The outcome that
would occur in a measurement of basis $\mathcal{B}$ can then be
inferred from the outcome in basis $\Omega$ by the association that
$U$ makes between elements of $\mathcal{B}$ and $\Omega$.

One might suspect therefore that the ontic state spaces of
Aaronson's models consist of the discrete set of basis states
$\Omega\subset\mathcal{PH}$, so that $\Lambda=\Omega$. However the
basis states $\Omega$ do not suffice to give a complete description
of the ontic configuration of a system, and we in fact have,
$\Lambda=\Omega\times\mathcal{PH}$. A specification of the preferred
basis states from $\Omega$ must be supplemented by specifying the
system's quantum state. Thus the quantum states describing a system
play a dual role, defining epistemic distributions over the subset
of ontic states from $\Omega$ whilst also playing an ontic role
themselves.
The epistemic states of Aaronson's models take the form,
\begin{equation}
\mu\left(\omega_i,\phi|\psi\right)\:d\phi=\delta\left(\phi-\psi\right)|\langle\omega_i|\phi\rangle|^2\:d\phi.
\end{equation}

That $|\psi\rangle$ must also play an ontological role becomes clear
from the indicator functions implied by Aaronson's models. These are
determined by the elements of the model's stochastic matrices. For
example, suppose that one wishes to perform a measurement in a basis
$\mathcal{B}$ on a system in state $|\psi\rangle$. Then, recalling
Aaronson's construction, we should rotate $|\psi\rangle$ with the
unitary $U:\mathcal{B}\rightarrow\Omega$. The probability of
obtaining an outcome $|j\rangle\in\mathcal{B}$ given that the
initial ontic state from $\Omega$ was $\omega_i$ is simply given by
the $ji^{th}$ element of the stochastic matrix $\mathbb{S}(U,\phi)$
(where we use the subscript $j$ to denote the basis state from
$\Omega$ which leads us to infer an outcome
$|j\rangle\in\mathcal{B}$). Hence the indicator function associated
with outcome $|j\rangle\in\mathcal{B}$ (i.e. with the projector
$|j\rangle\langle{j}|$) is given by,
\begin{equation}
\xi\left(j|\omega_i,\phi\right)=\mathbb{S}(U,\phi)_{ji}.\label{xi_aaronson}
\end{equation}

Note that because we must implement a rotation $U$ in order to
perform our measurement in the preferred basis, and because the
stochastic matrix associated with such a rotation necessarily
depends on the quantum state $|\phi\rangle$, then the indicator
function is also dependent on the system's state as well as the
basis state from $\Omega$. Thus we see that the most complete
description that the model can give of measurement outcomes requires
specifying the system's quantum state, not just the particular
$\omega_i\in\Omega$. Therefore the quantum state itself must play an
ontological role\footnote{One might suggest that the system's state,
$|\psi\rangle$ need not take an ontological role, but since it
defines an epistemic distribution over the preferred basis $\Omega$,
then perhaps it only introduces an epistemic component to the
indicator functions, thus changing their statistics without playing
an ontic role. However, this is not possible as one can simply see
by noting that the \textit{amplitudes} of a state in some fixed
basis are not sufficient to completely parameterize its position in
Hilbert space, and so this kind of epistemic dependence of $\xi$ on
$|\psi\rangle$ would not confer enough information about
$|\psi\rangle$ to allow $\xi$ to fully reproduce the quantum
statistics.}. These choices for epistemic states and indicator
functions reproduce the quantum statistics as required,
\begin{eqnarray}
\int\!{d}\Lambda\:\mu(\lambda|\psi)\:\xi(j|\lambda)\!&=&\!\sum_i\!\int{d}\phi\:\mu(\omega_i,\phi|\psi)\:\xi(j|\omega_i,\phi)\nonumber\\
&=&\!\sum_i\!\int{d}\phi\:\delta(\phi-\psi)|\langle\omega_i|\phi\rangle|^2\mathbb{S}(U,\phi)_{ji}\nonumber\\
&=&\!\sum_i\mathbb{S}(U,\psi)_{ji}\left|\langle\omega_i|\psi\rangle\right|^2\nonumber\\
&=&\!|\langle{j}|\psi\rangle|^2.
\end{eqnarray}
Where in the last line we have used the constraint on $\mathbb{S}$
given in (\ref{Aaronson_qmstat}) and the fact that
$U:\mathcal{B}\rightarrow\Omega$.

Eq.~(\ref{xi_aaronson}) shows that in Aaronson's models, the
indicator functions are dependent on the preparation procedure
$S_{\mathcal{P}}$ (i.e. what quantum state a system is prepared in).
However, this is not as pathological as one might suppose, since (as
was also the case in the Beltrametti-Bugajski model) the whole
preparation procedure $S_{\mathcal{P}}$ has an ontological status.
Thus the dependence of $\mathcal{M}$ on $S_{\mathcal{P}}$ is
directly mediated through the ontic states of the system. In
Sec.~\ref{SEC:ont_meas_prep}, we generalize the ontological
formalism in a way that can describe models in which indicator
functions have a dependence on $S_{\mathcal{P}}$ that cannot be
explained so simply.

It should also be noted that the ontic state space of Aaronson's
models is that of the Beltrametti-Bugajski model supplemented with
the preferred basis $\Omega$. Clearly, access to ontic states from
the Beltrametti-Bugajski model will not increase one's computational
power beyond that possible with standard quantum mechanics. It is
intriguing then that Aaronson is able to show in \cite{scott} that
models incorporating $\Omega$ as well as the Beltrametti-Bugajski
state space can offer increased computational power.



\subsubsection{Bell's first model \label{SEC:example_Bell1}}

In the paper preempting his famous theorem \cite{Bell_context}, J.
Bell described a very simple and (by his own admission) artificial
way of introducing `hidden variables' so as to reproduce the
predictions of quantum mechanics for a spin-$\tfrac{1}{2}$ system.
The model he introduced is outcome deterministic and valid for
quantum systems described by Hilbert spaces of any dimensionality.

The ontic states $\Lambda$ of Bell's first model can be written as
the cartesian product of two subspaces,
$\Lambda=\Lambda^{\prime}\times \Lambda^{\prime\prime}$. The first
of these subspaces is isomorphic to the projective Hilbert space of
the system in question, $\Lambda^{\prime}=\mathcal{PH}$, whilst the
second subspace is the unit interval
$\Lambda^{\prime\prime}=\left[0,1\right]$. A system prepared
according to a quantum state $|\psi\rangle$ is described in the Bell
model by an epistemic state that is separable over
$\Lambda^{\prime}$ and $\Lambda^{\prime\prime}$,
\begin{equation}
\mu(\lambda^{\prime},\lambda^{\prime\prime}|\psi)\:d\lambda^{\prime}d\lambda^{\prime\prime}=\mu(\lambda
^{\prime}|\psi)\mu(\lambda^{\prime\prime}|\psi)\:d\lambda^{\prime}d\lambda^{\prime\prime}.
\end{equation}
The distribution over $\Lambda^{\prime}$ picks out the relevant
$\lambda_{\psi}^{\prime}\in\Lambda^{\prime}$ corresponding to
$|\psi\rangle$;
$\mu(\lambda^{\prime}|\psi)=\delta(\lambda^{\prime}-\lambda_{\psi}^{\prime})d\lambda^{\prime}$,
whilst the distribution over $\Lambda^{\prime\prime}$ selects a
$\lambda^{\prime\prime}$ according to a uniform probability
distribution, regardless of the system's quantum state;
$\mu(\lambda^{\prime\prime}|\psi)=d\lambda^{\prime\prime}$. Thus the
epistemic state over whole ontic state space $\Lambda$ reads,
\begin{equation}
\mu(\lambda^{\prime},\lambda^{\prime\prime}|\psi)\:d\lambda^{\prime}d\lambda^{\prime\prime}=\delta(\lambda^{\prime}-\lambda_{\psi}^{\prime})\:d\lambda^{\prime}d\lambda^{\prime\prime}.
\end{equation}

Now suppose that we wished to perform an $N$ outcome PVM measurement
$P$, described in quantum mechanics by the projectors
$\left\{P_i\right\}_{i=1}^N$. Suppose furthermore that the system
has been prepared in a state $|\psi\rangle$. The ontic state of the
system will then be given by the pair
$(\lambda_{\psi}^{\prime},\lambda^{\prime\prime})$ (with
$\lambda^{\prime\prime}$ uniformly selected from the unit interval).
The model reproduces quantum statistics by partitioning the unit
interval, $\Lambda^{\prime\prime}$, into $N$ subsets, such that for
every $i\in\{1,\ldots,N\}$ a fraction
$\textrm{tr}(P_i|\psi\rangle\langle\psi|)$ of
$\lambda^{\prime\prime}\in\Lambda^{\prime\prime}$ are taken to yield
a positive outcome for $P_i$. Quantitatively then, Bell's first
model associates a deterministic indicator function with the
$i^{th}$ outcome which takes the form,

\begin{equation}
\xi(i|\lambda^{\prime},\lambda^{\prime\prime},P)=\Theta(\lambda^{\prime\prime}-x_{i-1}(\lambda^{\prime}))-\Theta(\lambda^{\prime\prime}-x_i(\lambda^{\prime})).
\end{equation}

Where the values $x_i(\lambda^{\prime})$ (determining the
$\lambda^{\prime\prime}$ over which
$\xi(i|\lambda^{\prime},\lambda^{\prime\prime},P)$ has support) are
given by,
\begin{equation}
x_0(\lambda^{\prime}_{\psi})=0,
\end{equation}
and,
\begin{equation}
x_i(\lambda^{\prime}_{\psi})=\sum_{j=1}^i\textrm{tr}(P_i|\psi\rangle\langle\psi|),
\end{equation}

\noindent for all other values of $i$. This gives precisely the
partitioning of the unit interval that we require. Note that we
assume some ordering of PVM elements is chosen for every
measurement, so that permuting the label, $i$, of the
$\left\{P_i\right\}_{i=1}^N$ does not change the indicator functions
associated with the projectors.

This model easily reproduces the quantum statistics for performing a
projective measurement $P_{\phi}=|\phi\rangle\langle\phi|$ on a
system prepared in state $|\psi\rangle$,

\begin{eqnarray}
\int{d}\lambda\:\mu(\lambda|\psi)\xi(\phi|\lambda)&=&\int{d}\lambda^{\prime}d\lambda^{\prime\prime}\delta(\lambda^{\prime}-\lambda_{\psi})\xi(\phi|\lambda^{\prime},\lambda^{\prime\prime})\nonumber\\
&=&\int{d}\lambda^{\prime\prime}\Theta(\lambda^{\prime\prime}-x_i(\lambda^{\prime}))\nonumber\\
&&-\int{d}\lambda^{\prime\prime}\Theta(\lambda^{\prime\prime}-x_{i-1}(\lambda^{\prime}))\nonumber\\
&=&|\langle\phi|\psi\rangle|^2.
\end{eqnarray}

\subsubsection{Bell's second model \label{SEC:example_Bell2}}

Bell also published a second hidden variable theory for
spin-$\tfrac{1}{2}$ systems, which was presented in the same paper
as his famous theorem \cite{Bell_locality}. As was the case in his
first model, two subsets of ontic states are employed in Bell's
second model, so again we write
$\Lambda=\Lambda^{\prime}\times\Lambda^{\prime\prime}$. This time
however, the first set of ontic states, $\Lambda^{\prime}$, are
taken as isomorphic to the set of points on the unit sphere. Thus
any given $\lambda^{\prime}\in\Lambda^{\prime}$ can be represented
by a unit vector, $\vec{\lambda}^{\prime}$. However, we will very
shortly see that as in the case of Aaronson's model, the indicator
functions of Bell's second model are dependent on the quantum state
a system is prepared in. Therefore, a complete description of the
system also requires a specification of a system's quantum state.
The second set of ontic states, $\Lambda^{\prime\prime}$, is hence
also isomorphic to the set of points on the unit sphere (since we
only consider spin-$\tfrac{1}{2}$ systems, this is equivalent to
taking $\Lambda^{\prime\prime}=\mathcal{PH}$). A spin-$\tfrac{1}{2}$
system prepared with its spin oriented along a direction $\vec{p}$
is then taken to be described by a pair
$(\vec{\lambda}^{\prime},\vec{\lambda}^{\prime\prime})$, where
$\vec{\lambda}^{\prime\prime}=\vec{p}$, and
$\vec{\lambda}^{\prime}\in\Lambda^{\prime}$ is chosen to lie, with
equal probability, at some point in the hemisphere of
$\Lambda^{\prime}$ defined by $\vec{p}$. Thus a preparation with
$S_{\mathcal{P}}=\vec{p}$ is described by an epistemic state over
$\Lambda$ of,
\begin{equation}
\mu(\vec{\lambda}^{\prime},\vec{\lambda}^{\prime\prime}|\vec{p})\:d\vec{\lambda}^{\prime}d\vec{\lambda}^{\prime\prime}=\frac{1}{2\pi}\delta(\vec{\lambda}^{\prime\prime}-\vec{p})\:\Theta(\vec{\lambda}^{\prime}\cdot\vec{\lambda}^{\prime\prime})\:d\vec{\lambda}^{\prime}d\vec{\lambda}^{\prime\prime}.
\label{Bell_epistemic}
\end{equation}

Now consider performing a measurement for whether or not the
system's spin lies along a direction $\vec{a}$. Bell's second model
specifies that we receive a positive outcome if the system's ontic
state $\vec{\lambda}^{\prime}\in\Lambda^{\prime}$ happens to lie in
the hemisphere centered on a vector $\vec{a}^{\prime}$. The vector
$\vec{a}^{\prime}$ is obtained by rotating the system's ontic state
$\vec{\lambda}^{\prime\prime}$ towards $\vec{a}$ through an angle
$\frac{\pi}{2}(1-\vec{\lambda}^{\prime\prime}\cdot\vec{a})$. Thus
the indicator function for a measurement of spin up along direction
$\vec{a}$ is given by,
\begin{equation}
\xi(+\vec{a}|\vec{\lambda}^{\prime},\vec{\lambda}^{\prime\prime})=\Theta(\vec{\lambda}^{\prime}\cdot\vec{a}^{\prime}),
\end{equation}
\noindent the dependence on $\vec{\lambda}^{\prime\prime}$ being
implicit within $\vec{a}^{\prime}$. This model reproduces the
required spin-$\tfrac{1}{2}$ quantum statistics as we would expect,
\begin{widetext}
\begin{eqnarray}
\int{d}\vec{\lambda}^{\prime}d\vec{\lambda}^{\prime\prime}\mu(\vec{\lambda}^{\prime},\vec{\lambda^{\prime\prime}}|\vec{p})\xi(+\vec{a}|\vec{\lambda}^{\prime},\vec{\lambda}^{\prime\prime})&=&\frac{1}{2\pi}\int{d}\vec{\lambda}^{\prime}d\vec{\lambda}^{\prime\prime}\:\delta(\vec{\lambda}^{\prime\prime}-\vec{p})\Theta(\vec{\lambda}^{\prime}\cdot\vec{\lambda}^{\prime\prime})\Theta(\vec{\lambda}^{\prime}\cdot\vec{a}^{\prime})\nonumber\\
&=&\frac{1}{2\pi}\int{d}\vec{\lambda}^{\prime}\:\Theta(\vec{\lambda}^{\prime}\cdot\vec{p})\Theta(\vec{\lambda}^{\prime}\cdot\vec{a}^{\prime})\nonumber\\
&=&\int_{\theta=0}^{\pi-\theta_{pa^{\prime}}}\int_{\phi=0}^{\pi}\sin{\theta}\:d\theta{d}\phi\nonumber\\
&=&\cos^2\frac{\theta_{pa^{\prime}}}{2}.
\end{eqnarray}
\end{widetext}

Where $(\theta,\phi)$ are polar coordinates and
$\theta_{pa^{\prime}}$ is the angle separating the unit vectors
$\vec{p}$ and $\vec{a}^{\prime}$.

As it stands, Bell's second model can be comfortably expressed in
the standard ontological model formalism. However, a slightly
modified version of this simple model shows the limitation of the
traditional formalism. In the above model the probabilistic nature
of the quantum statistics derives from an uncertainty in the
preparation of a system's ontic state (as can be seen from
Eq.~(\ref{Bell_epistemic})). But it is also possible to reformulate
the model so as to move this epistemic uncertainty into a lack of
knowledge of how the \textit{measuring device} is configured. Such a
possibility cannot be conceived of within the traditional
ontological model formalism, which only postulates ontic states for
the system. Clearly one needs to extend the formalism to include new
ontic states $\gamma_{\mathcal{M}}$ from a new ontic state space
$\Gamma_{\mathcal{M}}$ that act as a complete physical description
of $\mathcal{M}$. We will show in Sec.~\ref{SEC:ont_meas_prep} how
one can introduce such an extension whilst still reproducing quantum
mechanical predictions.

It is also worth noting that, as in Aaronson's models, the
measurement devices in both of Bell's models exhibit a dependence on
the preparation device's setting, but a dependence that is mediated
through the system's ontic state.

\subsubsection{Aerts' model \label{SEC:example_Aertmodel}}

Our final example model is the strongest motivation for the
extension we outline in the next section. Aerts has studied
ontological models which are entirely incompatible with the standard
ontological model formalism, since they explicitly treat the
measurement device at an ontological level. We will consider the
model given by Aerts in \cite{Aerts85,Aerts91} for
spin-$\tfrac{1}{2}$ quantum systems. This model attempts to
reproduce the quantum statistics through a rule for distributing
small spheres of charge on a unit sphere\footnote{In some instances,
Aerts presents his model using a sphere with non-unit radius,
although this is an unnecessary generalization for our purposes.}.
The preparation of a system $\mathcal{S}$ having spin along a
direction $\vec{r}$ is represented in the model by the placement of
a small sphere carrying a fixed positive charge $+q$ at point
$\vec{r}$ on the unit sphere. The charge $+q$ is in fact arbitrary,
and therefore a complete description of $\mathcal{S}$ is given by
$\vec{r}$ alone. Thus the ontic state space $\Lambda$ of the system
is isomorphic to the set of points on the surface of the unit sphere
and we can write the ontic state pertaining to $\mathcal{S}$ as
$\vec{\lambda}\in\Lambda$. If the preparation device $\mathcal{P}$
prepares a pure quantum state with Bloch vector $\vec{\psi}$, then
the epistemic state describing $\mathcal{S}$ will be,
\begin{equation}
\mu(\vec{\lambda}|\vec{\psi})\:d\vec{\lambda}=\delta(\vec{\lambda}-\vec{\psi})\:d\vec{\lambda}.
\label{Aerts_epistemic}
\end{equation}

A measurement of the system's spin along an arbitrary direction
$\vec{a}$ is represented by placing another two small spheres at
positions $\pm\vec{a}$, joined by a straight rigid rod passing
through the origin. These two spheres are also charged with negative
charges $-s$ and $-(1-s)$, where $s\in\left[0,1\right]$. The
particular value of $s$ is assumed to be unknown to the
experimentalist. These two charges, joined by a rigid rod,
constitute the measurement device, $\mathcal{M}$, of the model.
Given this arrangement Aerts specifies that the outcome of a spin
measurement is determined by which of the two charged spheres at
$\pm\vec{a}$ exerts the greater force on $+q$, consequently
attracting it. If the charge $+q$ ends up moving towards the sphere
at position $\vec{a}$ then an outcome of spin-up along $\vec{a}$ is
declared. If however, $+q$ ends up being attracted to the sphere at
$-\vec{a}$ then `spin-down' is announced.

Now note that there is no epistemic uncertainty in the ontological
configuration of the \textit{system} $\mathcal{S}$; if one knows
$S_{\mathcal{P}}$ then one also knows the ontic state (see
(\ref{Aerts_epistemic})). Therefore, as in the Beltrametti-Bugajski
model, the model must implement indeterministic indicator functions
over $\Lambda$ that directly mimic the quantum statistics that one
expects when measuring the spin along $\vec{a}$ of a system prepared
according to $\vec{\psi}$,

\begin{equation}
\xi(\vec{a}|\vec{\lambda})=\cos^2\frac{\theta_{a\lambda}}{2}.
\label{Aerts_indicator}
\end{equation}

Where $\theta_{a\lambda}$ is the angle between the vectors
$\vec{\lambda}$ and $\vec{a}$.

The epistemic states and indicator functions of Aerts' model take
essentially the same form as those from the Beltrametti-Bugajski
model and thus one might be tempted to see the two models as
equivalent. However the models differ crucially in how they treat
$\mathcal{M}$. The Beltrametti-Bugajski model does not specify the
nature of the indeterminism appearing in its indicator functions.
Aerts' model meanwhile, exhibits a specific construction for how
these probabilities could arise from an epistemic uncertainty of the
configuration, $s$, of $\mathcal{M}$.

It is clear from the description we have already given that Aerts'
model provides more structure to the operation of $\mathcal{M}$,
structure that we need in order to be able to distinguish it from
the Beltrametti-Bugajski model. To quantify this structure we will
need to create the extension of the ontological model formalism that
we also found lacking in our discussion of the Bell model. We now
finally present this extension, which will also allow us to view
contextuality as a restriction on interactions between the ontic
configurations of $\mathcal{S}$ and $\mathcal{M}$ (or $\mathcal{S}$
and $\mathcal{P}$ in the case of preparation contextuality). In
Sec.~\ref{SEC:ont_meas_prep} we will return to Aerts' model in
detail, distinguishing between two different possible manifestations
of outcome determinism which we term micro and macro-determinism
(originally alluded to in Sec.~\ref{SEC:om_intro}).

\section{Ontological treatment of measurement and preparation devices \label{SEC:ont_meas_prep}}


Our discussion so far has been greatly simplified by considering
preparation and measurement devices as external objects not
thoroughly treated by the theory, much as in an operational view of
quantum mechanics. This approach of only associating an ontic state
space $\Lambda$ with the system $\mathcal{S}$, has been the
traditional approach for discussing ontological models. Let us now
suppose that we provide $\mathcal{P}$ and $\mathcal{M}$ (which,
after all, are also physical systems) with the same ontological
treatment as $\mathcal{S}$, by introducing two new sets of ontic
states, $\gamma_{\mathcal{P}}\in\Lambda_{\mathcal{P}}$ and
$\gamma_{\mathcal{M}}\in\Lambda_{\mathcal{M}}$. The ontic states
from these sets describe the complete configurations of
$\mathcal{P}$ and $\mathcal{M}$ respectively.

Recall that the settings $S_{\mathcal{P}}$ and $S_{\mathcal{M}}$
denote configurations of the devices when they are set to perform
certain preparation or measurement procedures. There are many
different ontological configurations of $\mathcal{M}$ that we could
imagine being consistent with it still performing the same
measurement and thus being set according to the same
$S_{\mathcal{M}}$. For example, if we simply changed the color of
the paint on $\mathcal{M}$ then its ontological configuration -
being its \textit{complete} description - would change, but of
course the measurement it performs, and thus the setting
$S_{\mathcal{M}}$ describing it, would not be expected to change.
Therefore we can think of settings of $\mathcal{P}$ and
$\mathcal{M}$ as defining subsets of their ontic state spaces. We
denote the equivalence classes of ontic states consistent with
settings $S_{\mathcal{P}}$ and $S_{\mathcal{M}}$ (possibly
implemented according to some context\footnote{The Definitions
\ref{DEF:context} and \ref{DEF:prepcontext} that we shortly give for
preparation and measurement contexts show that, strictly speaking,
\textit{any} change of the ontic configuration
$\gamma_{\mathcal{M}}$ of $\mathcal{M}$ corresponds to a change of
measurement context. Thus even apparently trivial changes, such as
the color of the paint on $\mathcal{M}$, actually constitute
different measurement contexts. However, we will be interested in
changes of $\mathcal{M}$'s configuration that allow one to prove
measurement contextuality, and in general such contexts will
correspond to macroscopic alterations of $\mathcal{M}$'s state.
Generally then, a measurement context will define a subset of ontic
states contained within the set $\tilde{S}_{\mathcal{M}}$ associated
with a given setting $S_{\mathcal{M}}$.}) by
$\tilde{S}_{\mathcal{P}}\subset\Lambda_{\mathcal{P}}$ and
$\tilde{S}_{\mathcal{M}}\subset\Lambda_{\mathcal{M}}$. One thing
worth noting about this idea of `setting subsets' is that subsets
corresponding to different settings of either $\mathcal{P}$ or
$\mathcal{M}$ will necessarily be disjoint;
$\tilde{S}_{\mathcal{M}}\cap{\tilde{S}_{\mathcal{M}}^{\prime}}=\emptyset$
and
$\tilde{S}_{\mathcal{P}}\cap{\tilde{S}_{\mathcal{P}}^{\prime}}=\emptyset$
for $S_{\mathcal{M}}\neq{S}_{\mathcal{M}}^{\prime}$ and
$S_{\mathcal{P}}\neq{S}_{\mathcal{P}}^{\prime}$. This should be true
since knowledge of the ontic state of a device (being a complete
specification of its realistic description) allows us to completely
and \textit{uniquely} infer the device's setting.

How do preparation and measurement procedures on a system
$\mathcal{S}$ appear in terms of $\Gamma_{\mathcal{P}}$ and
$\Gamma_{\mathcal{M}}$? Performing a measurement on $\mathcal{S}$
involves an interaction between the ontic states of $\mathcal{S}$
and $\mathcal{M}$, an interaction ultimately allowing an observer to
infer pre-measurement information about the ontic state of
$\mathcal{S}$ from some macroscopic property of $\mathcal{M}$.
Similarly, a preparation of $\mathcal{S}$ corresponds to an
interaction between ontological configurations of $\mathcal{S}$ and
$\mathcal{P}$. Clearly then, the occurrence of measurement and
preparation procedures in an ontological model are crucially
dependent on how the model relates $\Lambda_{\mathcal{P}}$,
$\Lambda$ and $\Lambda_{\mathcal{M}}$. In order to be clear about
what assumptions we make about such relations we will begin with a
very general picture - one in which the three ontic state spaces do
not even individually exist - and gradually refine it by applying
appropriate assumptions on how they can interact. Eventually we
arrive at a formalism in which the ontological role of $\mathcal{M}$
(and $\mathcal{P}$) within the standard formalism from
Sec.~\ref{SEC:om_intro} is clear.

The most general possible description of $\mathcal{P}$,
$\mathcal{S}$ and $\mathcal{M}$ is one in which the three systems
are represented by a single non-separable reality, so that we cannot
even talk about individual systems $\mathcal{P}$, $\mathcal{S}$,
$\mathcal{M}$ or their individual ontic state spaces. Then the best
we can do is to speak of a single `global' ontic state space
$\Gamma$, containing ontic states $\nu$ which describe a
configuration of the whole $\mathcal{P},\mathcal{S},\mathcal{M}$
scenario. We then have epistemic states
$\mu(\nu|S_{\mathcal{P}},S_{\mathcal{M}})$ encoding the probability
of preparing a particular $\nu\in\Gamma$ given some settings
$S_{\mathcal{P}}$ and $S_{\mathcal{M}}$ of $\mathcal{P}$ and
$\mathcal{M}$. Similarly the indicator functions $\xi(j|\nu)$ in
such a non-separable model denote the probability of obtaining some
outcome $j$ of a measurement corresponding to setting
$S_{\mathcal{M}}$ given a particular $\nu$. The statistical
predictions of such a model are given by;
\begin{equation}
\text{Pr}(j|S_{\mathcal{P}},S_{\mathcal{M}})=\int\!\!{d}\nu\:\mu(\nu|S_{\mathcal{P}},S_{\mathcal{M}})\xi(j|\nu).
\end{equation}

Note that we do not write $\xi(j|\nu)$ as depending on
$S_{\mathcal{M}}$ since here we are allowing for the more general
case, where the indicator function not only depends on the
\textit{set} of ontic states defined by a setting $S_{\mathcal{M}}$,
but potentially on \textit{individual} ontic states themselves -
albeit non-separable ones, $\nu$.

In such non-separable models it is hard to build any intuitive
picture of reality whatsoever, with even the concepts of system,
preparation and measurement devices making little sense\footnote{See
\cite{Howard_einst_short} and \cite{Howard_einst_long} for a
discussion of the history of non-separability in realistic
interpretations of quantum mechanics.}. Consequently, all existing
models assume a \textit{separable} picture of reality for
$\mathcal{P}$, $\mathcal{S}$ and $\mathcal{M}$. This amounts to the
assumption that the `global' ontic state space of the three systems
can be written as a cartesian product of ontic state spaces for each
individual system,
$\Gamma=\Lambda_{\mathcal{P}}\times\Lambda\times\Lambda_{\mathcal{M}}$,
so that $\nu=(\gamma_{\mathcal{P}},\lambda,\gamma_{\mathcal{M}})$.
Models employing this assumption are constrained to reproduce
quantum statistics according to,
\begin{equation}
\text{Pr}(j|S_{\mathcal{P}},S_{\mathcal{M}})=\int_{\mathcal{P},\mathcal{S},\mathcal{M}}\!\!\!\!\!\!\!\!\mu(\gamma_{\mathcal{P}},\lambda,\gamma_{\mathcal{M}}|S_{\mathcal{P}},S_{\mathcal{M}})\xi(j|\gamma_{\mathcal{M}},\lambda,\gamma_{\mathcal{P}}).
\label{seperable}
\end{equation}

\noindent Where we adopt the shorthand
$\int_{\mathcal{P},\mathcal{S},\mathcal{M}}=\int\!\!\!\int\!\!\!\int{d}\gamma_{\mathcal{P}}{d}\lambda{d}\gamma_{\mathcal{M}}$.

The model thus now employs epistemic states
$\mu(\gamma_{\mathcal{P}},\lambda,\gamma_{\mathcal{M}}|S_{\mathcal{P}},S_{\mathcal{M}})$
and indicator functions
$\xi(j|\gamma_{\mathcal{P}},\lambda,\gamma_{\mathcal{M}})$ which
treat $\mathcal{P}$, $\mathcal{S}$ and $\mathcal{M}$ as having
separate ontic states. Thus we have arrived at a formalism
incorporating models in which indicator functions are dependent on
the settings of both the \textit{preparation} and measurement
devices. This formalism allows for cases where, unlike Bell's second
model and those considered by Aaronson, a dependence on
$S_{\mathcal{P}}$ is not simply mediated through the ontic states of
$\mathcal{S}$. In fact Eq.~(\ref{seperable}) can describe cases of
even greater generality, wherein measurement outcomes are dependent
on individual ontic states of $\mathcal{P}$ and $\mathcal{M}$, not
just the \textit{sets} of ontic states defined by their settings.

Eq.~(\ref{seperable}) employs single joint distributions over the
ontic states from all three systems, implicitly allowing for the
possibility that there is a \textit{statistical} dependence between
the ontic states of each system. There are a few reasonable
assumptions that we can make about the statistical relations that
might exist between the systems. The validity of these assumptions
can ultimately be called into question, but in fact our motivation
for using the formalism is precisely so that we can study the ways
in which such assumptions may \textit{fail} to do justice to our
universe. The hope is that we can pinpoint precisely which
assumptions are the troublemakers.

In most models, the configuration of a preparation device is taken
to only indirectly affect the outcome of any measurement via its
influence on the system $\mathcal{S}$. Our second assumption (after
separability), is therefore a statistical independence between
$\mathcal{M}$ and $\mathcal{P}$. Then not only are the ontic
configurations of the two devices independent of each other, but
furthermore the outcome of a measurement exhibits no \textit{direct}
statistical dependence on the preparation device's ontic state - any
such dependence having to be mediated through $\mathcal{S}$. Under
this assumption, Eq.~(\ref{seperable}) becomes,
\begin{eqnarray}
\text{Pr}(j|S_{\mathcal{P}},S_{\mathcal{M}})&=&\int_{\mathcal{P},\mathcal{S},\mathcal{M}}\!\!\!\!\!\!\!\!\mu(\gamma_{\mathcal{P}},\lambda,\gamma_{\mathcal{M}}|S_{\mathcal{P}},S_{\mathcal{M}})\xi(j|\lambda,\gamma_{\mathcal{M}}) \nonumber\\
&=&\int_{\mathcal{S},\mathcal{M}}\!\!\!\!\!\!\mu(\lambda,\gamma_{\mathcal{M}}|S_{\mathcal{P}},S_{\mathcal{M}})\xi(j|\lambda,\gamma_{\mathcal{M}}).
 \label{seperable2}
\end{eqnarray}

Where in the second line we have marginalized over the dependence on
the $\gamma_{\mathcal{P}}$, which (given our most recent assumption)
only appeared within the epistemic state
$\mu(\gamma_{\mathcal{P}},\lambda,\gamma_{\mathcal{M}}|S_{\mathcal{P}},S_{\mathcal{M}})$.

Although we can also consider an ontological treatment of
$\mathcal{P}$, for brevity we will now focus our attention solely on
the measurement device. To this end we can use an identity of
probabilities to write
$\mu(\lambda,\gamma_{\mathcal{M}}|S_{\mathcal{P}},S_{\mathcal{M}})=\mu(\gamma_{\mathcal{M}}|S_{\mathcal{P}},S_{\mathcal{M}})\mu(\lambda|S_{\mathcal{P}},S_{\mathcal{M}})$,
allowing us to further simplify (\ref{seperable2}) to,
\begin{equation}
\text{Pr}(j|S_{\mathcal{P}},S_{\mathcal{M}})=\int_{\mathcal{S},\mathcal{M}}\!\!\!\!\!\!\mu(\gamma_{\mathcal{M}}|S_{\mathcal{M}})\mu(\lambda|S_{\mathcal{P}},S_{\mathcal{M}})\xi(j|\lambda,\gamma_{\mathcal{M}}).
\label{seperable3}
\end{equation}

Where we have again used our assumption of statistical independence
of $\mathcal{P}$ and $\mathcal{M}$ to write
$\mu(\gamma_{\mathcal{M}}|S_{\mathcal{P}},S_{\mathcal{M}})=\mu(\gamma_{\mathcal{M}}|S_{\mathcal{M}})$.
Note that the epistemic state
$\mu(\lambda|S_{\mathcal{P}},S_{\mathcal{M}})$ allows the
$\lambda\in\Lambda$ to depend on the setting $S_{\mathcal{M}}$ of
$\mathcal{M}$. This kind of dependence is a formal expression of
what will introduce in Sec.~\ref{SEC:contextuality_interp} as
`$\lambda$-contextuality' - one of the possible ways of implementing
the kind of contextuality required by the Kochen Specker theorem
within the ontological model formalism. For the kind of models that
we consider, we make the explicit assumption that this kind of
dependence does not occur (as we justify in
Sec.~\ref{SEC:contextuality_interp}), so that which
$\lambda\in\Lambda$ applies to $\mathcal{S}$ is not dependent on the
ontic state $\gamma_{\mathcal{M}}$ describing $\mathcal{M}$.
Enforcing this assumption we therefore obtain,
\begin{equation}
\text{Pr}(j|S_{\mathcal{P}},S_{\mathcal{M}})=\int_{\mathcal{S},\mathcal{M}}\!\!\!\!\!\!\mu(\gamma_{\mathcal{M}}|S_{\mathcal{M}})\mu(\lambda|S_{\mathcal{P}})\xi(j|\lambda,\gamma_{\mathcal{M}}).
\label{seperable4}
\end{equation}

This is precisely the form that we need in order make it clear how
the traditional formalism can be adapted to provide an ontological
model for the \textit{measurement device} as well as the system.
Given knowledge of the measurement setting $S_{\mathcal{M}}$
describing $\mathcal{M}$, we obtain a distribution
$\mu(\gamma_{\mathcal{M}}|S_{\mathcal{M}})$ over its ontic states.
The particular ontic state describing $\mathcal{M}$, along with
$\lambda\in\Lambda$, then determines the outcome it produces - as is
clear from the form of the indicator function
$\xi(j|\lambda,\gamma_{\mathcal{M}})$. This formalism allows us to
describe ontological models such as that of Aerts, which provide a
more thorough realistic treatment of $\mathcal{M}$. We explicitly
show how Aerts' model can be expressed according to
(\ref{seperable4}) in the next section.


The expression in (\ref{seperable4}) thus shows how the standard
ontological model formalism would look if it were furnished with an
ontological model for $\mathcal{M}$. We can return to our completely
standard formalism (as introduced in Sec.~\ref{SEC:om_intro}) by
making one final assumption; that the measurement outcome depends
only on the measurement \textit{setting} of $\mathcal{M}$ and not on
the particular ontic state $\gamma_{\mathcal{M}}$. We can employ
this assumption by marginalizing the indicator function over
$\gamma_{\mathcal{M}}\in{S}_{\mathcal{M}}$, to give a
`coarse-grained' distribution, $\tilde{\xi}$,
\begin{equation}
\tilde{\xi}(j|\lambda,S_{\mathcal{M}})=\int_{\gamma_{\mathcal{M}}\in{\tilde{S}}_{\mathcal{M}}}\!\!d\gamma_{\mathcal{M}}\:\:\xi(j|\lambda,\gamma_{\mathcal{M}})\:\mu(\gamma_{\mathcal{M}}|S_{\mathcal{M}}).\label{marginalize_gammam}
\end{equation}
In doing this we are essentially eliminating the need for a model of
$\mathcal{M}$. Eq.~(\ref{seperable4}) then becomes,
\begin{equation}
\text{Pr}(j|S_{\mathcal{P}},S_{\mathcal{M}})=\int_{\mathcal{S}}\:\mu(\lambda|S_{\mathcal{P}})\tilde{\xi}(j|\lambda,S_{\mathcal{M}}).
\label{seperable5}
\end{equation}

Which is precisely our original formalism, as first introduced in
(\ref{ont_mod_qm_stats}). Clearly the implicit assumptions in this
standard formalism, highlighted in our above derivation, leave it
unable to describe a significant class of models, including those of
Aerts and the adapted version of Bell's second model.

Note that although here we have focused on showing how a measurement
device can be furnished with an ontological treatment, it is clear
that we can provide an ontological treatment of the preparation
device in an exactly analogous manner. This would lead us to
introduce a set of ontic states $\gamma_P\in\Gamma_P$ and an
epistemic distribution, $\mu(\gamma_P|S_{\mathcal{P}})$ describing
our knowledge of the ontic configuration of $\mathcal{P}$ given that
it is configured according to a setting $S_{\mathcal{P}}$.

\subsection{Models that measure with uncertainty \label{SEC:epistemic_meas_device}}

Eq.~(\ref{seperable4}) is exactly what we need to completely
describe Aerts' model, which we found ourselves ill-equipped to deal
with in Sec.~\ref{SEC:example_Aertmodel}.

Recall that Aerts' model aims to reproduce measurements made on a
spin-$\tfrac{1}{2}$ system, representing a measurement of spin along
direction $\vec{a}$ by spheres with negative charges of magnitudes
$s$ and $1-s$ lying at points $\pm\vec{a}$ on the unit sphere and
being connected by a rigid rod. The value of $s$ is chosen uniformly
at random from the interval $\left[0,1\right]$. Further recall that
a system prepared according to $|\psi\rangle$ is measured as having
spin-up (spin-down) along $\vec{a}$ if the net Coulomb force on a
sphere with charge $+q$, located at point $\vec{\psi}$ on the unit
sphere, attracts it towards the negatively charged sphere located at
$\vec{a}$ ($-\vec{a}$). The epistemic states and indicator functions
of Aerts' model are as given in (\ref{Aerts_epistemic}) and
(\ref{Aerts_indicator}). The key difference between the model of
Beltrametti-Bugajski and that of Aerts lies in the way that Aerts'
model treats the measurement device, since it introduces an ontic
state space for $\mathcal{M}$. The ontological configuration of
$\mathcal{M}$ consists of a specification of the arrangement of
negatively charged spheres constituting the device. To completely
specify this arrangement requires stating the orientation of the rod
holding the spheres and the value of $s\in\left[0,1\right]$
determining the charge held by the spheres. Thus the ontic state
space of $\mathcal{M}$ consists of two subspaces;
$\Gamma_{\mathcal{M}}=\Gamma_{\mathcal{M}}^{(1)}\times\Gamma_{\mathcal{M}}^{(2)}$,
and we write the respective ontic states as
$\vec{\gamma}_{\mathcal{M}}\in\Gamma_{\mathcal{M}}^{(1)}$ and
$s\in\Gamma_{\mathcal{M}}^{(2)}$ so that
$\gamma_{\mathcal{M}}\in\Gamma_{\mathcal{M}}$ is written as
$\gamma_{\mathcal{M}}=(\vec{\gamma}_{\mathcal{M}},s)$. The first
subspace, $\Gamma_{\mathcal{M}}^{(1)}$, is isomorphic to the unit
sphere, and $\vec{\gamma}_{\mathcal{M}}$ is simply taken to be the
vector $\vec{a}$ defining the rod's orientation. The second
subspace, $\Gamma_{\mathcal{M}}^{(2)}$, is given by the unit
interval, with $s$ being the charge on one of the spheres.

Now in Aerts' model, it is assumed that the value of $s$, although
it takes some definite value, is not known by the
experimenter\footnote{Aerts actually suggests a physical reason for
this within the context of his model, but this is not of importance
here.}. Thus there is an epistemic uncertainty with respect to the
precise configuration of $\mathcal{M}$. Therefore, following the
formalism of this section, we introduce an epistemic state
$\mu(\gamma_{\mathcal{M}}|S_{\mathcal{M}})$ describing the
configuration of the measurement device. Since the measurement
setting $S_{\mathcal{M}}$ of $\mathcal{M}$ is given by the direction
$\vec{a}$ (along which we wish to measure the system's spin) and $s$
is taken to be drawn uniformly at random from the interval $[0,1]$,
we have that,
\begin{equation}
\mu(\gamma_{\mathcal{M}}|S_{\mathcal{M}})\:d\gamma_{\mathcal{M}}=\delta(\vec{\gamma}_{\mathcal{M}}-S_{\mathcal{M}})\:d\vec{\gamma}_{\mathcal{M}}\:d{s}.
\end{equation}

To complete the ontological description of $\mathcal{M}$ we need an
indicator function specifying the outcome that $\mathcal{M}$ will
produce for given ontic states of $\mathcal{S}$ and $\mathcal{M}$
(of course the production of an `outcome' by $\mathcal{M}$ is
actually a certain evolution of $\mathcal{M}$'s ontic
configuration). In Aerts' model a measurement outcome is determined
by the relative strengths of the Coulomb attraction $F_{-a}$ (acting
on charge $+q$ at $\vec{\psi}$ due to the charge $-s$ located at
$-\vec{a}$) and the Coulomb attraction $F_a$ (due to the charge
$-(1-s)$ located at $\vec{a}$). Specifically, an outcome
corresponding to spin-up along $\vec{a}$ will occur if $F_a>F_{-a}$.
Using Coulomb's law, this requirement becomes \cite{Aerts91},
\begin{equation}
\frac{sq}{\pi\epsilon_0\sin^2(\theta_{a\psi}/2)}>\frac{(1-s)q}{\pi\epsilon_0\cos^2(\theta_{a\psi}/2)}.
\label{Aerts_force_inequality}
\end{equation}

Where we have denoted the angle separating the unit vectors
$\vec{a}$ and $\vec{\psi}$ as $\theta_{a\psi}$. According to
Eq.~(\ref{Aerts_force_inequality}), independently of $q$, an outcome
of spin up along $\vec{a}$ requires that we have
$s>\sin^2{\theta_{a\psi}/2}$. Therefore the indicator function
$\xi(+\vec{a}|\gamma_{\mathcal{M}},\lambda)$ (for the outcome
corresponding to measuring spin-up along direction $\vec{a}$) can be
written as,
\begin{eqnarray}
\xi(+\vec{a}|\gamma_{\mathcal{M}},\lambda)&=&\Theta(s-\sin^2{\tfrac{\theta_{a\psi}}{2}})\nonumber\\
&=&\Theta(s+\tfrac{1}{2}(\vec{\lambda}\cdot\vec{\gamma}_{\mathcal{M}}-1)).
\end{eqnarray}

Suppose we were to choose to coarse-grain over $\mathcal{M}$'s ontic
configuration, effectively ignoring any information we have about
its ontological model. Following (\ref{marginalize_gammam}) we
obtain an indicator function of the following form,
\begin{eqnarray}
\tilde{\xi}(+\vec{a}|\lambda,S_{\mathcal{M}})&=&\int{d}\gamma_{\mathcal{M}}\:\mu(\gamma_{\mathcal{M}}|S_{\mathcal{M}})\:\xi(+\vec{a}|\gamma_{\mathcal{M}},\lambda)\nonumber\\
&=&\int{d}\vec{\gamma}_{\mathcal{M}}d{s}\:\delta(\vec{\gamma}_{\mathcal{M}}-S_{\mathcal{M}})\:\xi(+\vec{a}|\gamma_{\mathcal{M}},\lambda)\nonumber\\
&=&\int{d}s\:\Theta(s-\sin^2{\tfrac{\theta_{a\psi}}{2}})\nonumber\\
&=&\cos^2\tfrac{\theta_{a\psi}}{2}.
\end{eqnarray}



This is precisely the `trivial' indicator function that we
attributed to Aerts' model in Sec.~\ref{SEC:example_Aertmodel}.

Aerts' model thus shows how introducing $\Gamma_{\mathcal{M}}$
allows us to reproduce quantum statistics through a lack of
knowledge of how measurements are implemented. In fact, Aerts' model
raises an interesting question about what outcome determinism really
means in models providing a full treatment of $\mathcal{M}$.

Previously we thought of an ontological model as being outcome
deterministic if it implemented idempotent indicator functions, so
that
$\xi^2(k|\lambda,S_{\mathcal{M}})=\xi(k|\lambda,S_{\mathcal{M}})\:\:\:\forall\:\lambda,k,S_{\mathcal{M}}$.
But in light of our previous discussion we now know that an
indicator function $\xi(k|\lambda,S_{\mathcal{M}})$ can actually
depend not just on the setting $S_{\mathcal{M}}$, but potentially on
the individual ontic states
$\gamma_{\mathcal{M}}\in{\tilde{S}}_{\mathcal{M}}$. One can
therefore consider classifying indicator functions by how they treat
individual ontic states of the \textit{measurement device}. Clearly
a deterministic indicator function must assign a \textit{constant}
value of either $0$ or $1$ to all $\gamma_{\mathcal{M}}$
corresponding to a certain measurement setting - i.e. all
$\gamma_{\mathcal{M}}\in{\tilde{S}}_{\mathcal{M}}$ must be treated
identically. In such a case, knowledge of the particular
$\gamma_{\mathcal{M}}\in{\tilde{S}}_{\mathcal{M}}$ pertaining to
$\mathcal{M}$ does not help one determine the outcome of a
measurement any better than simply knowing the setting
$S_{\mathcal{M}}$. We refer to a model which is outcome
deterministic in this manner as being \textit{macro}deterministic,

\begin{definition}
An ontological model is said to be \textbf{macrodeterministic} if
all measurement outcomes are determined given knowledge of the state
of a system and the macroscopic configuration of the measurement
device, $S_{\mathcal{M}}$. i.e,
\begin{equation}
\xi(j|\lambda,\gamma_{\mathcal{M}},S_{\mathcal{M}})=\xi^2(j|\lambda,\gamma_{\mathcal{M}},S_{\mathcal{M}}),
\end{equation}
and,
\begin{equation}
\xi(j|\lambda,\gamma_{\mathcal{M}},S_{\mathcal{M}})=\xi(j|\lambda,S_{\mathcal{M}})\:\:\:\forall\:\gamma_{\mathcal{M}}\in{\tilde{S}}_{\mathcal{M}}.
\end{equation}
\end{definition}

\noindent The idea being that measurement results in such outcome
deterministic models are \textit{macroscopically determined} by the
setting $S_{\mathcal{M}}$, being insensitive to the precise ontic
state of $\mathcal{M}$. It is of course alternatively possible that
the outcome of a measurement might be completely determined only if
we know the specific ontic state
$\gamma_{\mathcal{M}}\in\Gamma_{\mathcal{M}}$ of $\mathcal{M}$ as
well as $\lambda\in\Lambda$. In these models, specifying
$S_{\mathcal{M}}$ isn't enough, and measurement outcomes are
determined by the `microscopic' ontological configuration of
$\mathcal{M}$. Thus we term this class of models
\textit{micro}deterministic,

\begin{definition}
An ontological model is said to be \textbf{microdeterministic} if
the outcome of a measurement is not completely determined by
knowledge of the measurement setting $S_{\mathcal{M}}$ of a device
$\mathcal{M}$, but is furthermore dependent on the ontic
configuration $\gamma_{\mathcal{M}}\in{\tilde{S}}_{\mathcal{M}}$ of
$\mathcal{M}$. i.e.,
\begin{equation}
\xi(j|\lambda,\gamma_{\mathcal{M}},S_{\mathcal{M}})=\xi^2(j|\lambda,\gamma_{\mathcal{M}},S_{\mathcal{M}}),
\end{equation}
and,
\begin{equation}
\xi(j|\lambda,\gamma_{\mathcal{M}},S_{\mathcal{M}})\neq\xi(j|\lambda,\overline{\gamma}_{\mathcal{M}},S_{\mathcal{M}}),
\end{equation}
\noindent for some
$\gamma_{\mathcal{M}},\overline{\gamma}_{\mathcal{M}}\in{\tilde{S}}_{\mathcal{M}}$.
\end{definition}

\noindent Thus a microdeterministic model allows us to determine
definite outcomes for measurements so long as we know the precise
ontic configuration of the measuring device.

Now the interesting point to note \cite{Aerts85,coecke982} is that
microdeterministic models appear outcome \textit{in}deterministic if
we coarse-grain over $\Gamma_{\mathcal{M}}$. That is to say that if
measurement outcomes are dependent on the individual
$\gamma_{\mathcal{M}}\in\Gamma_{\mathcal{M}}$ but we are ignorant of
the exact value of $\gamma_{\mathcal{M}}$, then the best we can do
is assign probabilities for measurement outcomes based on our
restricted knowledge. If a model is microdeterministic then although
we may have
$\xi(j|\lambda,\gamma_{\mathcal{M}},S_{\mathcal{M}})\in\left\{0,1\right\}$,
the marginalized state $\tilde{\xi}(j|\lambda,S_{\mathcal{M}})$ can,
in general, only be expected to satisfy
$0\leq\tilde{\xi}(j|\lambda,S_{\mathcal{M}})\leq{1}$ (see
(\ref{marginalize_gammam})). This is illustrated nicely by Aerts'
model, which falls into the class of microdeterministic models.
Knowledge of $s$ is crucial in order to determine a measurement
outcome, and upon marginalizing
$\xi(j|\lambda,\gamma_{\mathcal{M}},S_{\mathcal{M}})$ over
$\gamma_{\mathcal{M}}\in{\tilde{S}}_{\mathcal{M}}$ we obtain an
\textit{indeterministic} indicator function.

Thus we see a mechanism by which a determinism - apparently inherent
as seen from the traditional ontological model formalism - can
actually arise from an \textit{epistemic} uncertainty regarding the
precise configuration of a measurement device. This possibility has
been investigated in rigorous mathematical detail by Coecke
\cite{coecke98,coecke982}.

\section{Contextuality}

So far we have developed a way of describing reality according to
ontological models, but that does little to tell us what
\textit{kind} of reality any particular ontological model might
describe. This information is expressed by the structure of its
ontic state space, $\Lambda$. Remarkably, there exist arguments
constraining the structure of \textit{any} realistic interpretation
of quantum mechanics (including ontological models) to possess
certain properties, such as nonlocality (Bell's theorem
\cite{Bell_locality}) and contextuality (the Kochen Specker theorem
\cite{Ks}). As described in the introduction, a key motivation for
studying ontological models is to identify such properties. Thus a
pertinent question is how known properties are manifested within the
ontological model formalism, a question which we address in this
section for the case of contextuality. Contextuality has been the
subject of much debate (see \cite{Bell_context} and
\cite{mermin_bell} for contrasting views) and $40$ years after its
inception it is still not clear what its necessity can teach us
about realism in quantum mechanics. After reviewing the idea of
contextuality we will use our extension to the ontological model
formalism (from Sec.~\ref{SEC:ont_meas_prep}) to show how it is
specifically manifested within these models. We are led to conclude
that contextuality, as it stands, can be implemented as a very
intuitive and unsurprising dynamical constraint. But the effect of
contextuality on ontological models can be more subtle, and in
Sec.~\ref{SEC:deficiency_intro} we will show how it implies a
property which we call deficiency. As we discuss in
Sec.~\ref{SEC:deficiency_interp}, deficiency prevents a natural
relationship between preparations and measurements in quantum
mechanics from being carried over to ontological models. We consider
this to be one case in which contextuality can quantitatively be
seen to give rise to unexpected behavior.

\subsection{What is Contextuality? \label{SEC:conextuality_intro}}

Contextuality has a long history, beginning in 1967, when Kochen and
Specker (KS) \cite{Ks} first introduced a notion which, following
\cite{spekkens_con}, we refer to as \textit{traditional
contextuality}\footnote{This has commonly been referred to simply as
contextuality, but we reserve this term for the more general notions
of contextuality that we introduce in Definitions
\ref{DEF:meas_contextuality_om} and \ref{DEF:prep_contextuality_om}
(originally introduced in \cite{spekkens_con}).} (TC). Consider
performing a projective measurement $|\psi\rangle\langle\psi|$ on a
system. In a two dimensional Hilbert space such a projector can be
uniquely implemented by a measurement procedure with outcomes
corresponding to $|\psi\rangle$ and $|\psi^{\perp}\rangle$ (where
$\langle\psi|\psi^{\perp}\rangle=0$). However, in a Hilbert space
with dimension greater than two, there is no unique way to
physically implement such a projector onto a single quantum state
$|\psi\rangle$. In an $N$ dimensional Hilbert space ($N>2$) one
implements $|\psi\rangle\langle\psi|$ as part of an $N$ outcome PVM,
where each outcome corresponds to one of $N$ orthogonal basis
states. Since there are a continuum of $N$ dimensional bases
containing the vector $|\psi\rangle$, there exist a continuum of PVM
measurements that can realize the projector
$|\psi\rangle\langle\psi|$. KS refer to the different PVMs that
contain a given rank one projector $|\psi\rangle\langle\psi|$ as the
\textit{contexts} of that projector.

In any outcome \textit{deterministic} and realistic view of nature
(regardless of whether or not it can be formalized in terms of an
ontological model), a projector $P$ is at all times assigned a
definite outcome `value', $v(P)\in\left\{0,1\right\}$, even before
it is measured. KS considered the possibility that a realistic
outcome deterministic theory might have to `change its mind' about
whether a value $0$ or $1$ is associated with a projector $P$
dependent on which PVM is used to implement it. Such a dependence is
what we refer to as \textit{traditional contextuality};


\begin{definition}
An outcome deterministic ontological model is said to be
\textbf{traditionally contextual} (TC) if there exists at least one
projection operator, $P$, such that the pre-determined outcome
$v(P)$ associated with $P$ is dependent on which PVM is used to
implement it. \label{DEF:MOC}
\end{definition}

TC therefore tells us that specifying that a measurement device is
configured to measure a projector $P$ is not sufficient in order to
uniquely identify the `real' value assigned to the result of its
measurement. Rather we must specify the whole PVM that we would set
$\mathcal{M}$ to measure. Incredibly, KS managed to show that TC, so
defined, must be possessed by all outcome deterministic realistic
theories reproducing the (experimentally verified) predictions of
quantum mechanics. We reproduce their ingenious proof in Appendix
\ref{APP:KS}, translated into the language of the ontological model
formalism.

In one sense, KS's proof of TC is extremely general. Associating a
pre-existing value $v(P)$ to a projector $P$ is a requirement of any
realistic outcome \textit{deterministic} theory and therefore TC is
defined (and proven by KS to be necessary) for \textit{any} such
theory, not only those that can be expressed in the ontological
model formalism. There are however, a few shortcomings of TC.
Definition \ref{DEF:MOC} only applies to systems described in
quantum mechanics by a Hilbert space of dimension greater than or
equal to $3$. Furthermore, it applies only to outcome
\textit{deterministic} realistic theories. Yet as was emphasized by
Bell \cite{bell_beables} and discussed in Sec.~\ref{SEC:om_intro},
an assumption of outcome determinism is quite distinct from one of
realism.
Another shortcoming of TC is that changing the PVM implementing a
projector is not the only change of $\mathcal{M}$'s setting that
quantum mechanics predicts should leave measurement outcome
statistics unaltered. For example there are many different ways of
convexly decomposing elements of a given POVM
measurement\footnote{Although the term `convex decomposition' does
not have a unique usage in the literature, we will say that a POVM
$E^{(0)}=\{E^{(0)}_k\}_k$ can be convexly decomposed in terms of a
set of other POVMs $E^{(1)},E^{(2)},\ldots,E^{(N)}$, if each of its
effects can be written in the form
$E^{(0)}_k=\sum_{i=1}^Np_iE^{(i)}_k$ with $\{p_i\}_{i=1}^N$ forming
a valid probability distribution.}, each of which provides a
different experimental arrangement in which one could physically
measure the same POVM elements.

Thus there are several reasons why TC appears a somewhat restricted
notion of contextuality, and one is led to wonder whether it is
possible to generalize the idea. Such a generalization was provided
by Spekkens in \cite{spekkens_con}. To begin with, one can broaden
the definition of a measurement context \cite{spekkens_con},

\begin{definition}
The possible \textbf{contexts} of the outcome of a measurement
performed by device $\mathcal{M}$ are all those measurement settings
$S_{\mathcal{M}}$ which do not alter the frequency of the outcome
when the measurement is performed on any particular preparation of a
system $\mathcal{S}$. \label{DEF:context}
\end{definition}



Different measurement procedures in quantum theory will give the
same outcome statistics so long as they are all described by the
same POVM element. \textit{Any} different settings $S_{\mathcal{M}}$
resulting in an outcome being described in quantum mechanics by the
same POVM (although perhaps written in another form) are therefore,
according to our above definition, different \textit{contexts} of
that outcome. We have already mentioned measurement contexts
associated with different PVMs realizing a given projector, and
different convex decompositions of a POVM. By Definition
\ref{DEF:context} there are clearly innumerable other possible
contexts. The macroscopic nature of $\mathcal{M}$ ensures that there
are a multitude of degrees of freedom one can manipulate whilst
effectively leaving the measurement operation of the device
un-altered. Of course many of these contexts would be hard to
formally quantify, and we restrict our consideration to those
contexts that can be described in a meaningful manner.

We can use Definition \ref{DEF:context} to introduce a generalized
notion of \textit{measurement contextuality} for both outcome
deterministic and \textit{in}deterministic ontological models
\cite{spekkens_con},

\begin{definition}
An ontological model is said to be \textbf{measurement
non-contextual} if it only associates a single indicator function
$\xi(k|\lambda,E)$ with a given POVM element $E_k$, regardless of
its context. Conversely a model is said to be \textbf{measurement
contextual} if the indicator function that it assigns to $E_k$
depends on its context, i.e. if there exist
$S_{\mathcal{M}},S_{\mathcal{M}}^{\prime}$ such that
$\xi(k|\lambda,E,S_{\mathcal{M}})\neq\xi(k|\lambda,E,S_{\mathcal{M}}^{\prime})$
(with $S_{\mathcal{M}}$ and $S_{\mathcal{M}}^{\prime}$ representing
different measurement contexts of the POVM effect $E_k$).
\label{DEF:meas_contextuality_om}
\end{definition}

According to this new definition, measurement contextuality is a
non-equivalence of a model's mathematical representations of those
measurements which quantum mechanics treats as being operationally
identical. As we noted previously, one can conceive of many
different measurement contexts and an ontological model could
potentially exhibit measurement contextuality with respect to any of
them. Therefore we must take care to specify with respect to which
context we might consider measurement contextuality at any given
time. As shown in \cite{spekkens_con}, Kochen and Specker's TC is
now seen to be a special case of this generalized measurement
contextuality. Specifically, TC corresponds to `measurement
contextuality with respect to the choice of PVM' in models that
exhibit outcome determinism for projective measurements.

In fact, following \cite{spekkens_con}, we can widen our concept of
contextuality even further by adapting Definition \ref{DEF:context}
to apply to \textit{preparations}. We define a preparation context
as follows,

\begin{definition}
The possible \textbf{contexts of a preparation} performed by device
$\mathcal{P}$ are all those preparation settings $S_{\mathcal{P}}$
of $\mathcal{P}$ which prepare a system $\mathcal{S}$ in states all
yielding identical measurement statistics for any particular
measurement performed on them. \label{DEF:prepcontext}
\end{definition}

Preparations that are described in quantum theory by the same
density operator always yield the same measurement statistics. Thus
different settings $S_{\mathcal{P}}$ of $\mathcal{P}$ described in
quantum theory by the same density operator (albeit perhaps the same
density operator written in a different form) are \textit{contexts}
of that preparation. As was the case with measurement contexts,
there are many ways one could vary $S_{\mathcal{P}}$ without
altering the density operator describing the measurement. For
example, there are many different ways of convexly decomposing a
mixed state density operator $\rho$. Each of these provide a
distinctly different probabilistic preparation procedure realizing
$\rho$, but yet all result in the same statistical predictions for
any measurement. Thus different convex decompositions of a density
operator form different contexts of a preparation.

Definition \ref{DEF:prepcontext} puts us in a position to consider
the possibility of \textit{preparation contextuality} within
ontological models \cite{spekkens_con},
\
\begin{definition}
An ontological model is said to be \textbf{preparation
non-contextual} if it only associates a single epistemic state
$\mu(\lambda|\rho)$ with a given density operator, $\rho$,
regardless of the preparation context. Conversely a model is said to
be \textbf{preparation contextual} if the epistemic state that it
assigns to $\rho$ depends on its context, i.e. there exists
$S_{\mathcal{P}},S_{\mathcal{P}}^{\prime}$ such that
$\mu(\lambda|\rho,S_{\mathcal{P}})\neq\mu(\lambda|\rho,S_{\mathcal{P}}^{\prime})$
(where $S_{\mathcal{P}}$ and $S_{\mathcal{P}}^{\prime}$ represent
different preparation contexts that realize the density operator
$\rho$). \label{DEF:prep_contextuality_om}
\end{definition}

It should be noted that there are cases where these generalized
definitions of preparation and measurement contextuality are
genuinely independent of each other. The Beltrametti-Bugajski model
for example exhibits preparation contextuality with respect to the
convex decompositions of a mixed state, but does not exhibit
measurement contextuality in the generalized sense of Definition
\ref{DEF:meas_contextuality_om}. To see this, note that in the
Beltrametti-Bugajski model a convex decomposition
$\rho=\sum_ip_i|\psi_i\rangle\langle\psi_i|$ of a mixed state $\rho$
into a set of pure states corresponds to an epistemic state
$\mu(\lambda|\rho)=\sum_ip_i\delta(\lambda-\lambda_{\psi})$ (see
Lemma \ref{LEM:mu_convex_relation} in Appendix
\ref{APP:elementary_results} for a justification). Clearly then,
different convex decompositions of $\rho$ will give epistemic states
having different supports, since the elements of the decomposition
\textit{are} precisely the ontic states. Hence we have preparation
contextuality. Conversely, the model will never exhibit
\textit{measurement} contextuality since the indicator function it
associates with a measurement is formed directly from that
measurement's quantum mechanical statistical predictions. This
clearly implies, according to Definition \ref{DEF:context}, that the
Beltrametti-Bugajski indicator functions will remain unaltered under
any change of context.


For a more in-depth example of contextuality, we can consider the KS
model, first introduced in Sec.~\ref{SEC:example_ksmodel}. This
exhibits both preparation and measurement contextuality. Its
preparation contextuality is with respect to the different possible
convex decompositions of a mixed state. To see this, consider a
mixed state described by a density operator $\rho$ which can be
prepared by either of the following two convex decompositions,

\begin{eqnarray}
\rho&=&\frac{3}{4}|0\rangle\langle{0}|+\frac{1}{4}|1\rangle\langle{1}|\nonumber\\
&=&\frac{1}{2}|\tfrac{\pi}{8}\rangle\langle\tfrac{\pi}{8}|+\frac{1}{2}|-\tfrac{\pi}{8}\rangle\langle-\tfrac{\pi}{8}|.
\label{decomps_pi8}
\end{eqnarray}

\noindent Where
$|\pm\tfrac{\pi}{8}\rangle=\cos{\tfrac{\pi}{8}}|0\rangle\pm\sin{\tfrac{\pi}{8}}|1\rangle$.
Denote the preparation setting that implements the first of these
convex decompositions as $S_{\mathcal{P}}$, and that which
implements the second decomposition as $S_{\mathcal{P}}^{\prime}$.

Lemma \ref{LEM:mu_convex_relation} in Appendix
\ref{APP:elementary_results} shows that an ontological model is
constrained to employ epistemic states for each of these settings
that respect the convex structures in (\ref{decomps_pi8}),

\begin{eqnarray}
\mu(\lambda|\rho,S_{\mathcal{P}})&=&\frac{3}{4}\mu(\lambda|0)+\frac{1}{4}\mu(\lambda|1)\nonumber\\
\mu(\lambda|\rho,S_{\mathcal{P}}^{\prime})&=&\frac{1}{2}\mu(\lambda|\tfrac{\pi}{8})+\frac{1}{2}\mu(\lambda|-\tfrac{\pi}{8}).
\end{eqnarray}

Now recall that in the Kochen Specker model, the epistemic state
associated with a quantum state has a support equal to the
hemisphere defined by the quantum state's Bloch vector. These
hemispheres are such that,

\begin{eqnarray}
\text{Supp}(\mu(\lambda|\rho,S_{\mathcal{P}}))&=&\text{Supp}(\mu(\lambda|0))\cup\text{Supp}(\mu(\lambda|1))\nonumber\\
&=&\text{Supp}(\Theta(\vec{0}\cdot\vec{\lambda})+\Theta(\vec{1}\cdot\vec{\lambda}))\nonumber\\
&=&\Lambda, \label{example_supp_equal}
\end{eqnarray}

and,

\begin{eqnarray}
\text{Supp}(\mu(\lambda|\rho,S_{\mathcal{P}}^{\prime}))&=&\text{Supp}(\mu(\lambda|\tfrac{\pi}{8}))\cup\text{Supp}(\mu(\lambda|-\tfrac{\pi}{8}))\nonumber\\
&=&\text{Supp}(\Theta(\vec{\tfrac{\pi}{8}}\cdot\vec{\lambda})+\Theta(-\vec{\tfrac{\pi}{8}}\cdot\vec{\lambda}))\nonumber\\
&\subset&\Lambda. \label{example_supp_less}
\end{eqnarray}

Where $\vec{0}$, $\vec{1}$, $\vec{\tfrac{\pi}{8}}$ and
$-\vec{\tfrac{\pi}{8}}$ denote the Bloch vectors associated with the
states $|0\rangle$,$|1\rangle$,$|\tfrac{\pi}{8}\rangle$ and
$|-\tfrac{\pi}{8}\rangle$ respectively.

Thus
$\text{Supp}(\mu(\lambda|\rho,S_{\mathcal{P}}))\neq\text{Supp}(\mu(\lambda|\rho,S_{\mathcal{P}}^{\prime}))$,
and consequently, according to Definition
\ref{DEF:prep_contextuality_om}, the Kochen Specker model is
preparation contextual. More specifically, note that
(\ref{example_supp_equal}) and (\ref{example_supp_less}) imply that
there are cases wherein the model realizes this contextuality by
changing the \textit{support} of an epistemic state as the
preparation context changes.

Now consider measurement contextuality in the KS model. To begin
with, note that since the model is for a two dimensional Hilbert
space it cannot possibly exhibit TC (in fact this was Kochen and
Specker's motivation for presenting this model). However, the model
does display measurement contextuality with respect to convex
decompositions of a POVM. Furthermore, the KS model implements this
measurement contextuality by changing the \textit{support} of an
indicator function as the measurement context is altered. We can see
this by employing precisely the same kind of construction as we used
to show its preparation contextuality. Specifically, consider the
POVM $\left\{E_1,E_2\right\}$ where the POVM elements have a
computational basis matrix representation of,
\begin{equation}
E_1=\left[
\begin{array}{cc}
\frac{3}{4} & 0 \\
0 & \frac{1}{4}
\end{array}
\right],\:\:\:\:E_2=\left[
\begin{array}{cc}
\frac{1}{4} & 0 \\
0 & \frac{3}{4}
\end{array}
\right].
\end{equation}

In particular, consider the element $E_1$. Two possible ways in
which we can realize this in terms of projective measurements are,
\begin{eqnarray}
E_1&=&\frac{3}{4}|0\rangle\langle{0}|+\frac{1}{4}|1\rangle\langle{1}|\label{E1_fg2}\\
&=&\frac{1}{2}|\tfrac{\pi}{8}\rangle\langle\tfrac{\pi}{8}|+\frac{1}{2}|-\tfrac{\pi}{8}\rangle\langle{-\tfrac{\pi}{8}}|.\label{E1_fg1}
\end{eqnarray}
Eqs.~(\ref{E1_fg2}) and (\ref{E1_fg1}) describe two different ways
of performing a measurement for whether or not a system would yield
the POVM outcome $E_1$. Eq.~(\ref{E1_fg2}) corresponds to a
measurement procedure in which we perform the PVM
$\left\{|0\rangle\langle{0}|,|1\rangle\langle{1}|\right\}$, yielding
an outcome of either `$0$' or `$1$'. We then randomly choose,
according to the distribution
$\left\{\frac{3}{4},\frac{1}{4}\right\}$, whether either the `$0$'
or `$1$' outcome will lead us to declare a positive outcome for
$E_1$. The second decomposition, Eq.~(\ref{E1_fg1}), stipulates that
we perform a similar protocol only this time we measure
$\left\{|\tfrac{\pi}{8}\rangle\langle\tfrac{\pi}{8}|,|-\tfrac{\pi}{8}\rangle\langle{-}\tfrac{\pi}{8}|\right\}$
and select which outcome should give $E_1$ from a uniform
distribution. Denote the configurations of $\mathcal{M}$ that
realize (\ref{E1_fg2}) and (\ref{E1_fg1}) as $S_{\mathcal{M}}$ and
$S_{\mathcal{M}}^{\prime}$ respectively. Given the relations in
Eqs.~(\ref{E1_fg2}) and (\ref{E1_fg1}) the KS model is constrained
to employ indicator functions satisfying (see Lemma
\ref{LEM:indfn_convex_relation} in Appendix
\ref{APP:elementary_results}),

\begin{eqnarray}
\xi(E_1|\lambda,S_{\mathcal{M}})&=&\frac{3}{4}\xi(0|\lambda)+\frac{1}{4}\xi(1|\lambda),\nonumber\\
\xi(E_1|\lambda,S_{\mathcal{M}}^{\prime})&=&\frac{1}{2}\xi(\tfrac{\pi}{8}|\lambda)+\frac{1}{2}\xi(-\tfrac{\pi}{8}|\lambda).
\end{eqnarray}

In the KS model, the supports of indicator functions associated with
projective measurements are hemispheres defined by the Bloch vector
of the state onto which they project. As before, the hemispherical
supports of these distributions are such that,
\begin{eqnarray}
\text{Supp}(\xi(E_1|\lambda,S_{\mathcal{M}}))&=&\text{Supp}(\xi(0|\lambda))\cup\text{Supp}(\xi(1|\lambda))\nonumber\\
&=&\text{Supp}(\Theta(\vec{0}\cdot\vec{\lambda})+\Theta(\vec{1}\cdot\vec{\lambda}))\nonumber\\
&=&\Lambda,
\end{eqnarray}
and,

\begin{eqnarray}
\text{Supp}(\xi(E_1|\lambda,S_{\mathcal{M}}^{\prime}))&=&\text{Supp}(\xi(\tfrac{\pi}{8}|\lambda))\cup\text{Supp}(\xi(-\tfrac{\pi}{8}|\lambda))\nonumber\\
&=&\text{Supp}(\Theta(\vec{\tfrac{\pi}{8}}\cdot\vec{\lambda})+\Theta(-\vec{\tfrac{\pi}{8}}\cdot\vec{\lambda}))\nonumber\\
&\subset&\Lambda.
\end{eqnarray}

Thus
$\text{Supp}(\xi(E_1|\lambda,S_{\mathcal{M}}))\neq\text{Supp}(\xi(E_1|\lambda,S_{\mathcal{M}}^{\prime}))$,
and we see that in some cases the KS model implements measurement
contextuality by having an indicator function's \textit{support}
depend on the measurement context.

Having seen these examples, one might wonder to what extent
ontological models \textit{must} exhibit these generalized kinds of
contextuality. In fact Spekkens has shown in \cite{spekkens_con}
that any ontological model associating deterministic indicator
functions with projective measurements must exhibit measurement
contextuality with respect to different convex decompositions of a
POVM\footnote{Note that the Beltrametti-Bugajski model does not
exhibit any kind of measurement contextuality. This is not in
contradiction with either the KS proof or these proofs of
generalized contextuality, since the model employs indeterministic
indicator functions for projective measurements, rendering it
outside of the scope of all known contextuality proofs.}.
Furthermore, any model must exhibit preparation contextuality with
respect to different convex decompositions of a density operator.

Thus the epistemic states and indicator functions that an
ontological model associates with certain preparations and
measurement outcomes must change dependent on the context that
realizes them. It is worth noting that, although it was the case in
the KS model, such a dependence does not necessarily require the
\textit{supports} of epistemic states or indicator functions to
change. i.e. we may not necessarily have a change in \textit{which}
ontic states could have been prepared by a preparation or might
produce a given measurement outcome. Instead it could be that only
the non-zero \textit{probability assignments} are altered for some
context-independent set of ontic states. The case is, however, more
clear-cut within those ontological models that are outcome
deterministic. Then measurement contextuality requires that
indicator functions \textit{must} change their supports since
deterministic indicator functions only assume values of $0$ or $1$.
Any change in their assignments amounts to a change of support!

Although the kinds of measurement and preparation contextuality
introduced in Definitions \ref{DEF:meas_contextuality_om} and
\ref{DEF:prep_contextuality_om} are the only kinds of contextuality
typically considered, there is another interesting possibility.
Recall that Acarinoses's indicator functions are dependent on the
quantum state that a system is \textit{prepared} in, and therefore
are dependent on the preparation setting $S_{\mathcal{P}}$. These
models thus introduce the possibility of a strange kind of
contextuality in which the indicator function associated with a
\textit{measurement} is dependent on the setting used for a system's
\textit{preparation}. In Aaronson's model this kind of contextuality
is somewhat trivial, since $S_{\mathcal{P}}$ is in fact an ontic
state, so it is entirely natural for the indicator functions to be
dependent on the setting of $\mathcal{P}$. In fact, as was seen in
Sec.~\ref{SEC:ont_meas_prep}, most models implicitly assume a lack
of direct statistical dependence between $\mathcal{P}$ and
$\mathcal{M}$, so this strange contextuality will only ever apply to
a small subset of ontological models.

\subsection{What does contextuality mean? \label{SEC:contextuality_interp}}

In Definition \ref{DEF:MOC} we gave a mathematical definition of TC,
and in Definitions \ref{DEF:meas_contextuality_om} and
\ref{DEF:prep_contextuality_om} we generalized the notion to
preparation and measurement contextuality. But we still lack a clear
picture of exactly what it is that these ideas of contextuality
really mean in our ontological model formalism - what kind of
structure might they enforce on the ontic state space $\Lambda$ and
its dynamics? Understanding this is crucial for us to even begin to
judge to what extent contextuality, like nonlocality, goes against
our intuition and might fundamentally prohibit a realistic view of
the quantum world.

First consider TC in our ontological model formalism. It is clear
that, since the definition of this property relies crucially on
assigning definite values to projective measurements, it can only
make sense in outcome \textit{deterministic} ontological models. We
therefore temporarily restrict ourselves to such cases. But even
under a deterministic restriction our ontological model formalism
does not explicitly talk about `assigning outcomes' to measurements,
as Definition \ref{DEF:MOC} does. Rather our formalism employs
indicator functions; assigning outcomes dependent on the ontic state
of $\mathcal{S}$. How then can we import TC into our formalism?

There are two ways that TC can be manifested within an ontological
model. Consider a system $\mathcal{S}$, described in quantum
mechanics by a three dimensional Hilbert space. Suppose that this
system actually resides in an ontic state $\lambda$ and that we use
a device $\mathcal{M}$ to perform a projective measurement
$P_0=|0\rangle\langle{0}|$ on $\mathcal{S}$. Now consider two
settings $S_{\mathcal{M}}$ and $S_{\mathcal{M}}^{\prime}$ of
$\mathcal{M}$ that can realize this measurement, taken to
respectively correspond to the PVM contexts
$\left\{P_0,P_1,P_2\right\}$ and
$\left\{P_0,P_1^{\prime},P_2^{\prime}\right\}$. In an outcome
deterministic ontological model, what \textit{outcome} is assigned
to projector $P_0$ (what we might refer to as $v(P_0)$) is
determined by whether or not
$\lambda\in\textrm{Supp}\left(\xi(P_0|\lambda)\right)$. Hence
\textit{in order for the outcome assigned to $P_0$ to be dependent
on the PVM setting (as required by TC), our model must ensure that
the inclusion of $\lambda$ in
$\textrm{Supp}\left(\xi(P_0|\lambda)\right)$ is dependent on this
setting}. We explicitly derive this requirement in Appendix
\ref{APP:KS}, where we recreate the original KS argument for TC in
the ontological model language (associating a deterministic
indicator function $\xi(P|\lambda)$ with a projective measurement
$P$, as opposed to a `value assignment', $v(P)$). Clearly there are
two ways in which the inclusion of $\lambda$ in
$\text{Supp}(\xi(P_0|\lambda))$ could change; either by changing
$\text{Supp}(\xi(P_0|\lambda))$ or by changing $\lambda$. We can
classify ontological models according to which of these
possibilities they use to realize TC,

\begin{definition} An ontological model is said to be
\textbf{$\xi$-contextual} if it realizes traditional contextuality
by changing the support of an indicator function as the setting of
$\mathcal{M}$ changes,
$S_{\mathcal{M}}\rightarrow{S}_{\mathcal{M}}^{\prime}$.
\end{definition}

\begin{definition}
An ontological model is said to be \textbf{$\lambda$-contextual} if
it realizes traditional contextuality by changing the ontic state
associated with a system as the setting of $\mathcal{M}$ changes,
$S_{\mathcal{M}}\rightarrow{S}_{\mathcal{M}}^{\prime}$.
\end{definition}

In $\xi$-contextual models we have that a change of $\mathcal{M}$'s
setting simply changes the indicator function associated with
$\mathcal{M}$, thus changing how $\mathcal{M}$ will respond to
$\mathcal{S}$ during the measurement process. In
$\lambda$-contextual models however, a change of measurement setting
can result in the ontic configuration of the system $\mathcal{S}$
being changed - a potentially nonlocal effect if $\mathcal{S}$ and
$\mathcal{M}$ are space-like separated. Two models that differ only
in which of these approaches they use to realize TC are, according
to Definition \ref{DEF:ont_equiv}, ontologically equivalent. The
$\lambda$-contextual versus $\xi$-contextual distinction is thus a
purely metaphysical one; for any model implementing
$\lambda$-contextuality there is an entirely equivalent model that
implements $\xi$-contextuality. Bearing this in mind, we can justify
the assumption we made in Sec.~\ref{SEC:ont_meas_prep}; that
$\mu(\lambda|S_{\mathcal{P}},S_{\mathcal{M}})=\mu(\lambda|S_{\mathcal{P}})$.
Throughout the remainder of the paper we continue to assume that TC
is always implemented through $\xi$-contextuality.

Having discussed the manifestation of TC we now turn to the
generalized notions of preparation and measurement contextuality
given in Definitions \ref{DEF:meas_contextuality_om} and
\ref{DEF:prep_contextuality_om}. Understanding these types of
contextuality requires using the ontic state spaces
$\Gamma_{\mathcal{P}}$ and $\Gamma_{\mathcal{M}}$ for $\mathcal{P}$
and $\mathcal{M}$ that we outlined in the formalism derived in
Sec.~\ref{SEC:ont_meas_prep}.
To begin with, refer to Eq.~(\ref{seperable4}) from that section.
This shows explicitly that the measurement process within an
ontological model amounts to an interaction between the ontic states
of $\mathcal{S}$ and $\mathcal{M}$. Specifically, the indicator
function $\xi(j|\lambda,\gamma_{\mathcal{M}})$ tells us whether the
result of an interaction between a given $\lambda\in\Lambda$ and
$\gamma_{\mathcal{M}}\in{\tilde{S}_{\mathcal{M}}}$ would leave
$\mathcal{M}$ in a configuration such that we would infer the
$j^{th}$ measurement outcome to have occurred.
Now recall that a change of context implies a change of the device
setting $S_{\mathcal{M}}$ and consequently a change of
$\mathcal{M}$'s ontic state, $\gamma_{\mathcal{M}}$. Contextuality
requires that $\xi(j|\lambda,S_{\mathcal{M}})$ changes along with
$S_{\mathcal{M}}$. Thus contextuality actually imposes a restriction
on the interaction between $\mathcal{S}$ and $\mathcal{M}$. In
particular, the interaction - encoded within
$\xi(j|\lambda,S_{\mathcal{M}})$ - must change for any change of
$\gamma_{\mathcal{M}}$ that corresponds to an alteration of
measurement context.

The conclusion of this brief analysis, which can similarly be
performed for $\mathcal{P}$, is that the requirements of
contextuality can be satisfied by the completely natural arrangement
that the interaction of $\mathcal{M}$ and $\mathcal{S}$ be dependent
on the configuration of $\mathcal{M}$. A trivially simple example of
how contextuality can in principle be manifested in this natural way
can be found by introducing ontic states $\Gamma_{\mathcal{P}}$ and
$\Gamma_{\mathcal{M}}$ to the KS model from
Sec.~\ref{SEC:example_ksmodel}. In Sec.~\ref{SEC:conextuality_intro}
we saw how the KS model exhibits contextuality for convex
decompositions of POVMs by having an indicator function
$\xi(k|\lambda,E)$ change dependent on whether $E$ is performed
using a setting $S_{\mathcal{M}}$ in which either
$|0\rangle\langle{0}|$ or $|1\rangle\langle{1}|$ is measured or a
setting $S_{\mathcal{M}}^{\prime}$ in which either
$|\tfrac{\pi}{8}\rangle\langle\tfrac{\pi}{8}|$ or
$|-\tfrac{\pi}{8}\rangle\langle{-\tfrac{\pi}{8}}|$ is measured.
Suppose that we adopt an ontological model for measurement devices
within the KS model where $\gamma_{\mathcal{M}}$ is given precisely
by the Bloch vector associated with the projective measurement that
$\mathcal{M}$ is configured to perform. We can then explicitly see
that the two different settings of $\mathcal{M}$ correspond to
different ontological configurations of the measurement device;
$\vec{\gamma}_{\mathcal{M}}\in\{\vec{0},\vec{1}\}$ for setting
$S_{\mathcal{M}}$ and
$\vec{\gamma}_{\mathcal{M}}\in\{\vec{\tfrac{\pi}{8}},\vec{\tfrac{\pi}{8}}\}$
for setting $S_{\mathcal{M}}^{\prime}$. Thus contextuality simply
amounts to the measurement outcome being dependent on the
ontological condition of $\mathcal{M}$.

The discussion in this section suggests contextuality to be an
entirely natural requirement of realistic theories, in no way
comparable to the un-intuitive nature of nonlocality. This is an
intuition which was also held by Bell with regards to traditional
contextuality \cite{Bell_context},

\begin{quote}
``The result of an observation may reasonably depend not only on the
state of the system (including hidden variables) but also on the
complete disposition of the apparatus.''
\end{quote}

But despite what we have seen, it is possible that contextuality may
not be quite so simple to interpret. On analyzing the contextual
interactions between $\mathcal{S}$ and $\mathcal{M}$ in more detail
one might find stronger restrictions on their dynamics, restrictions
that may seem less natural than a simple dependence of a measurement
outcome on the interaction between $\mathcal{S}$ and $\mathcal{M}$.
The point we wish to emphasis however, is that as far as we are
aware, there do not exist proofs showing the necessity of such
stronger constraints\footnote{It should be noted though that there
is at least one exception to this statement. Consider tailoring the
measurement arrangement of $\mathcal{M}$ to be such that a change of
its measurement context corresponds to altering parts of
$\mathcal{M}$ that are space-like separated. Then the requirement of
non-contextuality actually becomes a requirement of nonlocality
\cite{mermin_bell}. However, this special case does not shed any
light on the implications of contextuality in situations where it
possesses an identity separate from nonlocality.}, and existing
proofs hint only towards the natural kind of dependence outlined
above.

In the next section we use contextuality to deduce a property that
must be possessed by ontological models: deficiency. Deficiency
tells us that the realistic states in an ontological model are
unable to respect certain operational relations between preparations
and measurements from quantum mechanics. We argue in
Sec.~\ref{SEC:deficiency_interp} that we would intuitively expect
these relations to carry over to a realistic description of quantum
mechanics, and thus deficiency is at least one aspect of
contextuality that demonstrates restrictions stronger than one might
have expected.

\section{Deficiency \label{SEC:deficiency_intro}}

An interesting feature of the KS model is that the epistemic state
associated with preparing a system according to state $|\psi\rangle$
has a support equal to the support of the indicator function
associated with performing a projective measurement
$|\psi\rangle\langle\psi|$ (see Eqs.~(\ref{mu_kochen_specker}) and
(\ref{xi_ks})). That is,
$\textrm{Supp}(\mu(\lambda|\psi))=\textrm{Supp}(\xi(\psi|\lambda))$.
This property is \textit{not} possessed however, by Bell's first
model. By considering how its epistemic states and indicator
functions act over the subset $\Lambda^{\prime}$ of ontic states we
can see that
$\textrm{Supp}(\mu(\lambda|\psi))\subset\textrm{Supp}(\xi(\psi|\lambda))$.
This lack of an equality between supports of the epistemic states
and indicator functions associated with preparing or measuring the
same quantum state $|\psi\rangle$ is what we call
\textit{deficiency}. Bell's first model is thus deficient, whilst
the KS model fails to exhibit the property. The KS model also fails
to exhibit TC, which it could not possibly exhibit since it is only
defined for two dimensional Hilbert spaces. Bell's first model
however, being an outcome deterministic model for Hilbert spaces
having dimension greater than $2$, is bound by the Kochen Specker
theorem to exhibit TC. One might therefore speculate at the
possibility of some kind of relationship between TC and deficiency.
In fact we will shortly show that any model exhibiting contextuality
for projective measurements (of which TC is the only known
quantified example\footnote{Note however, that one can envision
other (un-quantified) projective measurement contexts, such as the
possibility of altering a measurement's von Neumann chain.}) must
exhibit deficiency.

First however, we must take a moment to define deficiency more
rigorously. In the brief introduction presented above, we referred
to there being a single epistemic state $\mu(\lambda|\psi)$
associated with a preparation $|\psi\rangle$, and a single indicator
function $\xi(\psi|\lambda)$ associated with a projective
measurement $|\psi\rangle\langle\psi|$. The discussion in
Sec.~\ref{SEC:conextuality_intro} showed that in some cases one
cannot get away with only associating a single indicator function
with $|\psi\rangle\langle\psi|$. Rather, TC implies that to
unambiguously specify an indicator function one will also need to
specify the context of a measurement. It is also a possibility
(although it has not yet been proven to be a necessity) that more
than one epistemic state could be associated with a given
\textit{pure} state preparation $|\psi\rangle$, depending on the
setting $S_{\mathcal{P}}$ used to prepare it. Thus referring to
deficiency as meaning
$\text{Supp}(\mu(\lambda|\psi))\subset\text{Supp}(\xi(\psi|\lambda))$
is somewhat ambiguous. With respect to which contexts do we need
this expression to hold? Accordingly, we adopt a refined idea of
deficiency. An ontological model will be said to \textit{not} be
deficient if
$\text{Supp}(\mu(\lambda|\psi,S_{\mathcal{P}}))=\text{Supp}(\xi(\psi|\lambda,S_{\mathcal{M}}))$
for all ${S}_{\mathcal{P}},S_{\mathcal{M}}$, where
$S_{\mathcal{P}}$, and $S_{\mathcal{M}}$ denote full specifications
of the device's settings, including their context. Note that such an
equality between supports is the only possibility in a non-deficient
model, since we show in Lemma \ref{LEM:Lemma_mu_lt_xi} of Appendix
\ref{APP:elementary_results} that the epistemic states and indicator
functions of any model must always satisfy
$\text{Supp}(\mu(\lambda|\psi,S_{\mathcal{P}}))\subseteq\text{Supp}(\xi(\psi|\lambda,S_{\mathcal{M}}))$.
Thus we rigorously classify ontological models as deficient by the
following criteria,

\begin{definition}
An ontological model is said to be \textbf{deficient} if there
exists a pure quantum state $|\psi\rangle$ for which we have,
\begin{equation}
\text{Supp}(\mu(\lambda|\psi,S_{\mathcal{P}}))\subset\text{Supp}(\xi(\psi|\lambda,S_{\mathcal{M}})),
\end{equation}
for \textbf{some particular} $S_{\mathcal{P}}$ and
$S_{\mathcal{M}}$. \label{DEF:deficiency}
\end{definition}
%
In terms of the quantum formalism, deficiency states that the set of
ontic states possibly describing a system prepared in a quantum
state $|\psi\rangle$ cannot be the same as those ontic states
triggering a positive outcome for a measurement
$|\psi\rangle\langle\psi|$.

It is quite simple to show that TC implies deficiency, so that any
proof of the necessity of TC in ontological models implies the
necessity of deficiency as a simple corollary.

\begin{theorem}
Any ontological model capable of describing systems of dimension
greater than $2$ must be deficient in the sense of Definition
\ref{DEF:deficiency}. \label{THRM:deficiency}
\end{theorem}

\begin{proof}

The proof of this theorem proceeds in two parts. We first present a
simple argument showing that any outcome \textit{indeterministic}
ontological model must be deficient. Following this we complete the
proof by showing that deficiency must also apply to all outcome
deterministic models, so long as TC holds.

First then, consider outcome indeterministic models, and suppose for
a reductio ad absurdum, that deficiency does \textit{not} hold. Then
there would exist some quantum state preparation $|\psi\rangle$ and
associated projective measurement $|\psi\rangle\langle\psi|$ for
which the model employs epistemic states and indicator functions
satisfying,
\begin{equation}
\text{Supp}(\mu(\lambda|\psi,S_{\mathcal{P}}))=\text{Supp}(\xi(\psi|\lambda,S_{\mathcal{M}})),
\label{no_deficiency}
\end{equation}
\noindent for all $S_{\mathcal{P}}$ and $S_{\mathcal{M}}$.

Now since we expect that a system prepared in a state $|\psi\rangle$
should \textit{always} pass a projective measurement test
$|\psi\rangle\langle\psi|$ then we require,
\begin{equation}
\int{d}\lambda\:\:\mu(\lambda|\psi)\xi(\psi|\lambda)=1.
\label{psi_meas_psi}
\end{equation}

However, since $\mu(\lambda|\psi)$ is a normalized probability
distribution over $\Lambda$ (see Eq.~(\ref{mu_norm})) then we can
only satisfy (\ref{psi_meas_psi}) by having $\xi(\psi|\lambda)=1$
for all $\lambda\in\text{Supp}(\mu(\lambda|\psi))$. But if
deficiency does not hold then this would also imply that
$\xi(\psi|\lambda)=1$ for all
$\lambda\in\text{Supp}(\xi(\psi|\lambda))$ - i.e. that
$\xi(\psi|\lambda)$ is a \textit{deterministic} indicator function,
contrary to our initial assumption. Thus we conclude that if a model
is outcome indeterministic then it must be deficient.

Now we turn to outcome deterministic ontological models. For another
reductio ad absurdum, we again consider an ontological model that is
not deficient, so that again (\ref{no_deficiency}) holds. Now fix a
preparation setting $S_{\mathcal{P}}$. Eq.~(\ref{no_deficiency})
then implies that we will have,
\begin{equation}
\text{Supp}(\mu(\lambda|\psi,S_{\mathcal{P}}))=\text{Supp}(\xi(\psi|\lambda,S_{\mathcal{M}}))\:\:\forall\:{S}_{\mathcal{M}},
\end{equation}

and thus,
\begin{equation}
\text{Supp}(\xi(\psi|\lambda,S_{\mathcal{M}}))=\text{Supp}(\xi(\psi|\lambda,S_{\mathcal{M}}^{\prime})),
\label{nodef_violatecontext}
\end{equation}

\noindent for \textit{any} two measurement settings
$S_{\mathcal{M}}\neq{S}_{\mathcal{M}}^{\prime}$. But recalling
Definition \ref{DEF:meas_contextuality_om},
Eq.~(\ref{nodef_violatecontext}) has shown that,
\begin{equation}
\lnot\text{Deficiency}\:\implies\:\lnot\text{TC},
\end{equation}

and so,
\begin{equation}
\text{TC}\:\implies\:\text{Deficiency}.\label{MOC_imply_def}
\end{equation}

But we know from the Kochen Specker argument (reproduced in Appendix
\ref{APP:KS}) that there exists some $|\psi\rangle$ and some
settings $S_{\mathcal{M}}$, $S_{\mathcal{M}}^{\prime}$ for which TC
can be proven to occur in any outcome deterministic ontological
model of quantum mechanical systems having dimension greater than
$2$. Thus we deduce from (\ref{MOC_imply_def}) that any such outcome
deterministic ontological model must be deficient.

\end{proof}

We have trivially been able to show that any outcome indeterministic
model must be deficient. But the possibility remains that outcome
deterministic models of $2$ dimensional quantum systems may not be
deficient. This is because Theorem \ref{THRM:deficiency} shows that
deficiency results when deterministic indicator functions are
dependent on the measurement setting $S_{\mathcal{M}}$. This occurs
when a model exhibits TC, which it cannot if it describes a two
dimensional system. But deficiency can also follow if deterministic
indicator functions are dependent on the \textit{preparation}
setting of a system, $S_{\mathcal{P}}$. As we noted in Sections
\ref{SEC:example_Aaronsonmodel} and \ref{SEC:example_Bell2}, both
Aaronson's model and Bell's second model - through their choice of
ontic state space - exhibit a dependence of $\mathcal{M}$ on
$S_{\mathcal{P}}$, and indeed we can see that both these models are
deficient. For example, in the case of Bell's second model,
$\textrm{Supp}(\mu(\lambda|\vec{\psi}))=\mathfrak{H}(\vec{\psi})\times\{\vec{\psi}\}$
(where $\mathfrak{H}(\vec{\psi})$ is the hemisphere of
$\Lambda^{\prime}$ centered on $\vec{\psi}$), so that epistemic
states are restricted to have their supports over only one element
$\vec{\psi}\in\Lambda^{\prime\prime}$, determined by their
preparation setting $S_{\mathcal{P}}$. The indicator functions
however, due to their dependence on the system's quantum state (i.e.
$S_{\mathcal{P}}$), have the larger support
$\textrm{Supp}(\xi(+\vec{a}|\lambda))=\bigcup_{\vec{\lambda}^{\prime\prime}\in\Lambda^{\prime\prime}}\mathfrak{H}(\vec{a}^{\prime}(\vec{\lambda}^{\prime\prime}))\times\{\vec{\lambda}^{\prime\prime}\}$,
where we have written
$\vec{a}^{\prime}(\vec{\lambda}^{\prime\prime})$ to make clear the
implicit dependence of $\vec{a}^{\prime}$ on
$\vec{\lambda}^{\prime\prime}$ (see Sec.~\ref{SEC:example_Bell2}).
This support includes $\textrm{Supp}(\mu(\lambda|\vec{\psi}))$ as
the special case $\vec{\lambda}^{\prime\prime}=\vec{\psi}$, since
then $\vec{a}^{\prime}(\vec{\psi})=\vec{\psi}$. Thus deficiency is
achieved because of $\xi(\vec{a}|\lambda)$'s dependence on
$\Lambda^{\prime\prime}$, i.e. on $S_{\mathcal{P}}$.

One might also be led to think of deficiency as an implication of TC
(since for outcome deterministic models it is the existence of TC
that ensures deficiency). But we have also trivially managed to show
that deficiency must exist in outcome indeterministic ontological
models, for which TC cannot possibly be exhibited. Thus deficiency
actually holds for a wider class of ontological models than TC.

\subsection{Interpreting deficiency \label{SEC:deficiency_interp}}

We have seen that for a large class of ontological models the set of
ontic states possibly describing a system prepared in a quantum
state $|\psi\rangle$ cannot be the same as those ontic states
triggering a positive outcome for a measurement
$|\psi\rangle\langle\psi|$. But what implications does this
deficiency have? How would a deficient ontological model behave? We
now attempt, through an analogy, to describe the operational
implications of deficiency, and show that it is in some sense a
surprising property of ontological models. Of course arguments like
this that intend to address `intuitiveness' or `surprisingness' are
highly subjective, but regardless of how it is read, the analogy we
provide below nevertheless gives some picture of the behavior of
deficient ontological models.

Crucial to our definition of deficiency is that we implicitly hold
there to be an association between a quantum preparation
$|\psi\rangle$ and a quantum measurement $|\psi\rangle\langle\psi|$,
warranting a comparison of the associated epistemic states and
indicator functions. This is of course motivated by the fact that
$|\psi\rangle\langle\psi|$ is the unique rank one measurement for
which the Born rule yields an outcome probability of $1$ for a
system prepared according to $|\psi\rangle$. Understanding the role
of this association in an ontological model is key to understanding
deficiency. To this end, we digress into a simple analogy from
classical physics.

Imagine a toy system consisting of a small ball $b$, and suppose
that a complete description of the ball is given by a specification
of its position. The ball is a completely classical object, and so
it will always have some definite position regardless of whether or
not it is observed. Suppose that the possible preparation and
measurement procedures that one can perform on $b$ are defined by
boxes fixed at definite positions in space. Preparing $b$ `according
to a box $B_P$' implies that $b$ is known to reside at some definite
but unknown position within $B_P$ immediately after the time of
preparation. The boxes thus represent a restriction on our ability
to know the exact position at which $b$ is placed during a
preparation. Similarly suppose that the measurements that we can
perform on $b$ are restricted to being performed `according to some
box $B_M$'. By this we mean that the outcome of such a measurement
would tell us only whether or not the ball resided within that box,
but not its exact position. This `box-world' is in many ways
analogous to our ontological model constructions. The position of
$b$ - being a complete description of the ball system - is analogous
to our system ontic states $\lambda\in\Lambda$ and the boxes $B_P$
and $B_M$ are representative of the \textit{supports} of epistemic
states and indicator functions over $\Lambda$.

Now scientists living in box-world, perhaps through some perverse
historical accident, have come to adopt a theory that they refer to
as the `box-o-centric' theory. In this theory it is asserted that
\textit{there is no ball} $b$, and no concept of real positions at
all - only the abstract concept of a box\footnote{Scientists from
box-world who support the box-o-centric theory are likely to feel at
home with operational quantum mechanics.}. Such a theory, wherein we
talk only about boxes, is very strange, since, although the concept
of a box exists, there is no notion of `being contained' within a
box, and furthermore (since box-o-centrics do not find position to
their taste) no way in which to distinguish between boxes in terms
of the positions at which they reside.

We mentioned above how we associate a quantum preparation
$|\psi\rangle$ and projective measurement $|\psi\rangle\langle\psi|$
because of the probabilities obtained through the Born rule. Suppose
that a box-o-centric scientist wanted to similarly try and associate
box preparations and measurements. Now a box-o-centric advocate is
unable to compare the positions of two boxes as a reference for such
associations, since she does not believe in such concepts. Thus the
only way a box-o-centric could identify a measurement as being a
measurement \textit{of} box $B$ would be in a way analogous to how
we identify $|\psi\rangle\langle\psi|$ as a measurement of
$|\psi\rangle$ in quantum mechanics. That is, test whether that
measurement always gives a positive outcome when performed on
systems prepared according to box $B$. A box satisfying this
criterion would, as far as the scientist is concerned, be the best
candidate box for performing a measurement of box $B$. Thus
box-o-centrics are restricted to only compare boxes in an
operational fashion.

But box-o-centrics are not the only scientists in box-world, there
are also box-realists, who heretically believe that positions exist.
They propose that $b$ always resides somewhere, regardless of the
fact that box-world scientists are somehow condemned to only ever
possess incomplete information about its position. Now these
box-realists, who have no qualms with positions, would naturally
hope that preparation and measurement boxes which had previously
been identified with each other by box-o-centrics would
\textit{actually} be equal - i.e. enclose the same positions. Then
whenever a measurement of box $B$ had been performed, it really
would have been telling us that $b$ had been prepared with a
position inside the box $B$.

We can level a similar hope at quantum mechanics; that upon
introducing the idea of ontic states, a measurement
$|\psi\rangle\langle\psi|$ will remain associated with a preparation
$|\psi\rangle$. By this we mean that we would like the ontological
description to be such that any system yielding a positive outcome
for $|\psi\rangle\langle\psi|$ will have been described by an ontic
state that a preparation $|\psi\rangle$ could have left it in. i.e.
we would hope that
$\text{Supp}(\mu(\lambda|\psi))=\text{Supp}(\xi(\psi|\lambda))$.
Since $|\psi\rangle\langle\psi|$ is the best measurement that
quantum mechanics provides for testing whether a system is in a
state $|\psi\rangle$, our proof of deficiency shows that
\textit{there is no quantum measurement that would allow us to
deduce with certainty whether the \textbf{ontological configuration}
of a system was compatible with a preparation $|\psi\rangle$}. So
deficiency tells us that the real description introduced by an
ontological model cannot, as we might have hoped, maintain the
operational association we make between preparations and
measurements in quantum mechanics. Any such realistic theory must
instead exhibit a more complicated structure.

One can also use deficiency to restrict the dynamics an ontological
model implements upon measurement. For every state, there are
occasions when a measurement of the projector onto that state will
necessarily induce a disturbance of a system's ontic state. To help
show this, denote,
\begin{equation}
\mathcal{D}(\psi)=\textrm{Supp}(\xi(\psi|\lambda))-\textrm{Supp}(\mu(\lambda|\psi)).\label{Def_lambdas}
\end{equation}
Deficiency shows that there exist states $\psi$ such that
$\mathcal{D}(\psi)\neq\emptyset$. It will also be helpful to note
that for any $\lambda\in\mathcal{D}(\psi)$ one can always find a
state $|\phi\rangle$ such that $\lambda$ falls in the support of its
epistemic state. Furthermore, since its epistemic state has a finite
overlap with $\mathcal{D}(\psi)$, this state $|\phi\rangle$ will
neither be equal, nor orthogonal to $|\psi\rangle$.

Now the update rule that quantum mechanics specifies when obtaining
an outcome $\Pi_{\psi}=|\psi\rangle\langle\psi|$ of some PVM
performed on a system in state
$\rho_{\phi}=|\phi\rangle\langle\phi|$ is,
\begin{equation}
\rho_{\psi}=\frac{\Pi_{\psi}\rho_{\phi}\Pi_{\psi}}{\textrm{tr}(\Pi_{\psi}\rho_{\phi})}.
\end{equation}
One might expect that the analogous update rule in a deterministic
ontological model would be,
\begin{equation}
\mu(\lambda|\psi)=\frac{\mu(\lambda|\phi)\xi(\psi|\lambda)}{\int{d}\lambda\:\mu(\lambda|\phi)\xi(\psi|\lambda)}.\label{updaterule_om}
\end{equation}
Which essentially `projects' the system's epistemic state onto the
support of $\xi(\psi|\lambda)$. However, this non-disturbing update
rule must fail in deficient ontological models because deficiency
requires $\lambda$ to be disturbed upon measurement, as we now show.

Suppose that we implement the preparation of a system in some
quantum state by filtering the results of a measurement on the
system. For example, assuming a von Neumann collapse rule, a
preparation of a system $\mathcal{S}$ in state $|\psi\rangle$ can be
effected by performing a PVM measurement, $\mathbb{P}$, containing
the rank one projector $|\psi\rangle\langle\psi|$, on $\mathcal{S}$
and then post-selecting only those systems that yield the outcome
corresponding to $|\psi\rangle\langle\psi|$. The systems that will
survive this measure-and-filter procedure and be prepared in state
$|\psi\rangle$ are thus those yielding the outcome
$|\psi\rangle\langle\psi|$ of $\mathbb{P}$. Therefore any system
described by an ontic state satisfying
$\lambda\in\textrm{Supp}(\xi(\psi|\lambda))$ will successfully be
prepared in state $|\psi\rangle$ by this method.

Now naturally, we require that the ontic state of any system said to
be prepared in a state $|\psi\rangle$ should satisfy
$\lambda\in\textrm{Supp}(\mu(\lambda|\psi))$. But as we noted
previously, a system can be configured according to an ontic state
$\lambda\in\mathcal{D}(\psi)$, such that the measure-and-filter
procedure will prepare it in state $|\psi\rangle$, but yet it does
not satisfy the associated requirement
$\lambda\in\textrm{Supp}(\mu(\lambda|\psi))$. Measurements in a
deficient model must employ some kind of ontic state dynamics to
rectify this inconsistency, and consequently
Eq.~(\ref{updaterule_om}) cannot correctly describe the measurement
process in ontological models. Whilst one can give much simpler
proofs to show that realistic theories must provide such a
disturbance on measurement, our derivation shows explicitly how
disturbance can be related to the contextual nature of a theory.

\section{Conclusions\label{SEC:conclusions}}

We have outlined how one can increase the scope of an existing
formalism for realistic theories to include models that consider
measurement procedures in more detail. An often debated topic is
whether or not contextuality is a truly surprising requirement of
ontological models. In Sec.~\ref{SEC:contextuality_interp} we have
shown how our quantitative description of measurement devices allows
contextuality (to the extent that it is normally considered) to be
realized as a reasonable dynamical constraint on the interaction of
a system and measurement device. However, as we stressed previously,
this leaves open the possibility that these constraints might, under
further investigation, take on a more pathological form. Indeed, in
Sec.~\ref{SEC:deficiency_intro} we went some way in this direction
by arguing that deficiency - the fact that one cannot faithfully
associate measurements with a given preparation - provides at least
one aspect of contextuality which is manifestly not so reasonable.
If nothing else, we see this as evidence that there is more to be
said about contextuality, and that judgement of its implications
should be reserved until one can quantitatively analyze its effects
in more depth.

Addressing problems from quantum information using a realistic
approach to quantum mechanics can be a powerful tool, a fact
highlighted by Aaronson's work on complexity and hidden variables.
Using the ontological model formalism, we have characterized those
models to which his results apply. Another key motivation for our
study of ontological models is to quantify the conceptual problems
of quantum mechanics relative to a realistic framework. Crucially,
we see the utility of a realistic approach as being able to
highlight these problems in a familiar language, regardless of
whether the approach satisfactorily solves them. Essentially, our
motivation is to see what properties a realistic theory must possess
in order to reproduce quantum mechanics. In this paper we have tried
to build on this approach by going some way to quantitatively
clarifying the status of contextuality and introducing the property
of deficiency. Much work remains to be done in order to fully
understand the requirements of any realist theory reproducing
quantum mechanics, but there has been evidence \cite{toy_theory} to
show that this approach may indeed be a fruitful one.

\section{Acknowledgements}

We thank Robert Spekkens for many helpful discussions and for proof
reading this paper. NH acknowledges support from Imperial College
London. TR acknowledges support from EPSRC.

\begin{appendix}

\section{Proving KS theorem for ontological models \label{APP:KS}}

The argument that Kochen and Specker (KS) used \cite{Ks} to prove
that any realistic theory must exhibit traditional contextuality
(TC) involves a remarkable geometric construction which we now
outline. KS did not derive their proof using the formalism we
introduced in Sec.~\ref{SEC:om_intro}, but instead worked in a
simpler (yet more restrictive) formalism wherein they simply
consider assigning a definite outcome value $v(P)\in\{0,1\}$ to any
projective measurement $P$. We now review their proof within our
ontological model approach, allowing us to directly see traditional
contextuality's repercussions for these theories.

\begin{figure}[t]
\includegraphics[scale=0.4]{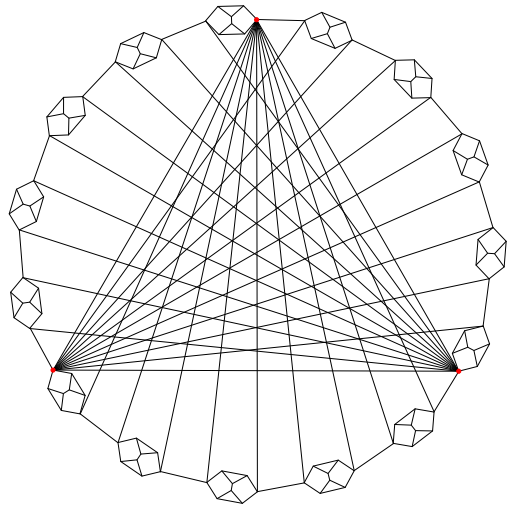}
\caption{The graph used by Kochen and Specker in \cite{Ks} to
provide a geometric impossibility argument proving the necessity of
traditional contextuality.} \label{FIG:KSgraph}
\end{figure}

The argument begins by considering a specific set,
$\Psi=\left\{|\psi_i\rangle\right\}_{i=1}^{116}$, of $116$ states
pertaining to a quantum system $\mathcal{S}$ described by a three
dimensional Hilbert space. KS represent these states graphically by
assigning a graph vertex to each of the $116$ states, and connecting
vertices by an edge if they correspond to orthogonal states. The
spectacular resulting graph, shown in Fig.~\ref{FIG:KSgraph}, can be
considered an `orthogonality map' of the set $\Psi$.

The elements of $\Psi$ are taken to define a set of $116$ projective
measurements that one could performed on a $3$ dimensional quantum
system,
$\mathbb{P}=\left\{|\psi_i\rangle\langle\psi_i|\right\}_{i=1}^{116}$.
An element $|\psi_i\rangle\langle\psi_i|\in\mathbb{P}$ represents a
test for whether a system is in the state $|\psi_i\rangle$. An
outcome deterministic ontological model introduces a set of $116$
indicator functions
$\Xi=\left\{\xi\left(i|\lambda\right)\right\}_{i=1}^{116}$, which
are distributions over the ontic state space of $\mathcal{S}$. These
are taken to be in one-to-one correspondence with the set
$\mathbb{P}$ of projectors. We restrict our attention, as KS did in
their original argument, to outcome deterministic ontological
models\footnote{The generalized notion of contextuality that we
introduce in Sec.~\ref{SEC:conextuality_intro} can be applied to
outcome indeterministic models.}. Therefore, when evaluated at some
fixed, but otherwise arbitrary ontic state
$\lambda^{\prime}\in\Lambda$, each element of $\Xi$ will specify an
assignment of either $0$ or $1$ to its corresponding projector from
$\mathbb{P}$. Equivalently, we can think of each element of $\Xi$ as
specifying, for each $\lambda^{\prime}\in\Lambda$, an assignment of
$0$ or $1$ to each vertex in Fig.~\ref{FIG:KSgraph}. The task at
hand for an ontological model is to perform these binary assignments
to graph vertices in a way consistent with the predictions of
quantum theory. KS re-word this task as a graph coloring problem by
representing the assignment of $0$ or $1$ to each element of
$\mathbb{P}$ as a coloring of the corresponding vertex in
Fig.~\ref{FIG:KSgraph} as either \textit{red} (for an assignment of
value $0$) or \textit{green} (for an assignment of value $1$). The
task faced by an ontological model is then to color the vertices of
Fig.~\ref{FIG:KSgraph} in a way consistent with quantum mechanics.
But just what are the restrictions that quantum theory imposes on
such a coloring? Having defined their coloring scheme, KS derive a
set of three \textit{coloring constraints}, which are imposed on any
coloring of Fig.~\ref{FIG:KSgraph} by the predictions of quantum
mechanics,

\begin{enumerate}
\item \textit{Every} vertex must be colored either red or green.
\item Any three vertices forming a triangle (a \textit{triad} of vertices) must be colored such
that one and only one is green.
\item Any two connected vertices must be colored so that both of
them are not green.
\begin{equation}
\label{colouring_constraints}
\end{equation}
\end{enumerate}

The first of these constraints is the defining requirement of
realism, whilst the second and third can be deduced from Lemmas
\ref{LEM:xi_pvm_span_lambda} and \ref{LEM:xi_pvm_disjoint} in
Appendix \ref{APP:elementary_results}. To see how, note that since
we consider a $3$ dimensional quantum system, vertices forming a
triad are associated with three mutually orthogonal states. Thus
each triad defines a PVM measurement on $\mathcal{S}$. Lemmas
\ref{LEM:xi_pvm_span_lambda} and \ref{LEM:xi_pvm_disjoint} together
imply that one and only one of the triad of (assumed
indeterministic) indicator functions associated with a PVM can
assign a value of $1$ for any given $\lambda\in\Lambda$. Thus one
and only one vertex from a triad can be colored green. Having given
a coloring scheme and a set of constraints on how the scheme must be
applied, KS then employ a geometrical argument to show that coloring
Fig.~\ref{FIG:KSgraph} according to conditions
(\ref{colouring_constraints}) is impossible (see
\cite{Ks,mermin_bell,redhead} for outlines of this geometrical
argument).

There is however, an implicit and subtle assumption required for
this proof to go through. Specifically, there is a subset of
vertices in Fig.~\ref{FIG:KSgraph} which belong to more than one
triad, and the geometrical part of the KS argument implicitly needs
to assume that one does not alter the color assigned to such a
vertex dependent on which triad it is considered to reside in. Since
every triad defines a PVM then such a dependence would constitute a
reliance of the \textit{outcome} assigned to a projector on the
particular PVM that a measurement device $\mathcal{M}$ employs to
measure it. This is precisely TC. Hence for the KS impossibility
argument to go through one must assume a realistic theory which
assigns values in a traditionally \textit{non}-contextual manner.
Thus Fig.~\ref{FIG:KSgraph} can be colored consistently only by
traditionally contextual theories. Consistently coloring
Fig.~\ref{FIG:KSgraph} is a mandatory requirement of any realistic
interpretation of quantum mechanics (since such a view should assign
all attributes pre-existing values) and so we are forced to conclude
that any realistic theory must exhibit TC.

It is worth mentioning that in the traditional formulation of the KS
argument there are several subtleties involved in assuming that
quantum mechanical statistics can be taken to imply constraints on
an realistic theories predictions for \textit{individual}
measurement outcomes (see \cite{redhead} for more details). These
philosophical subtleties still remain in our adaption of the
argument. In particular, they implicitly arise in our derivation of
Lemmas \ref{LEM:xi_pvm_span_lambda} and \ref{LEM:xi_pvm_disjoint}
which are crucial in deriving the coloring constraints
(\ref{colouring_constraints}).

\section{Elementary results \label{APP:elementary_results}}

There are several simple relations between the epistemic states and
indicator functions of any ontological model which can be seen to
follow almost immediately from the definitions of these
distributions. In this appendix we outline those relations which are
of use to us in the main text.

Firstly we note a simple relation between the supports of epistemic
states and indicator functions;

\begin{lemma}
The epistemic state $\mu(\lambda|\psi)$ associated with a
preparation of state $|\psi\rangle$ must have a support contained
entirely within the support of the indicator function
$\xi(\psi|\lambda)$ associated with a projective measurement
$|\psi\rangle\langle\psi|$,
\begin{equation}
\text{Supp}(\mu(\lambda|\psi))\subseteq\text{Supp}(\xi(\psi|\lambda)).
\label{musupp_less_xisupp}
\end{equation}
\label{LEM:Lemma_mu_lt_xi}
\end{lemma}

\begin{proof}
Quantum mechanical statistics tell us that a system prepared
according to $|\psi\rangle$ should \textit{always} pass a test
$|\psi\rangle\langle\psi|$, since $|\langle\psi|\psi\rangle|^2=1$.
An ontological model attempts to reproduce this result through the
integral,
\begin{eqnarray}
\int_{\Lambda}d\lambda\:\mu(\lambda|\psi)\xi(\psi|\lambda)&=&\int_{\text{Supp}(\xi(\psi|\lambda))}\!\!\!\!\!\!\!\!\!\!\!\!\!d\lambda\:\mu(\lambda|\psi)\xi(\psi|\lambda)\nonumber\\
&=&1.
\end{eqnarray}
Where we have made use of the fact that the integral over $\Lambda$
will only be non-zero within $\text{Supp}(\xi(\psi|\lambda))$.
Noting that $0\leq\xi(\psi|\lambda)\leq{1}$ and recalling the
normalization constraint (\ref{mu_norm}) on $\mu(\lambda|\psi)$ we
see that this integral can only take the required value of $1$ if it
includes the whole of $\text{Supp}(\mu(\lambda|\psi))$. Thus we
require (\ref{musupp_less_xisupp}) to hold.
\end{proof}
\\

Whether or not we can see the inclusion in
(\ref{musupp_less_xisupp}) to be \textit{strict} is the subject of
our discussion of deficiency in Sec.~\ref{SEC:ont_meas_prep}.

Lemma \ref{LEM:Lemma_mu_lt_xi} allows us to immediately deduce a
useful fact regarding the epistemic states associated with
orthogonal quantum states \cite{spekkens_con},

\begin{lemma}
Epistemic states $\mu(\lambda|\psi)$ and $\mu(\lambda|\psi^{\perp})$
associated with two orthogonal quantum states $|\psi\rangle$ and
$|\psi^{\perp}\rangle$ must have disjoint supports;
\begin{equation}
\text{Supp}(\mu(\lambda|\psi))\cap\text{Supp}(\mu(\lambda|\psi^{\perp}))=\emptyset.
\label{mu_orthog_disjoint}
\end{equation}
\end{lemma}

\begin{proof}
This result follows simply from noting that quantum mechanics
predicts that a state $|\psi\rangle$ should \textit{never} pass a
test for being in an orthogonal state $|\psi^{\perp}\rangle$
($|\langle\psi|\psi^{\perp}\rangle|^2=0$). In order for an
ontological model to respect this we clearly need to have that
$\text{Supp}(\mu(\lambda|\psi))\cap\text{Supp}(\xi(\psi^{\perp}|\lambda))=\emptyset$
(and
$\text{Supp}(\mu(\lambda|\psi^{\perp}))\cap\text{Supp}(\xi(\psi|\lambda))=\emptyset$),
since otherwise it is possible that a preparation of $|\psi\rangle$
could result in $\mathcal{S}$ being prepared in an ontic state that
could then trigger a positive outcome in a measurement of
$|\psi\rangle\langle\psi|$. Referring to Lemma
\ref{LEM:Lemma_mu_lt_xi} we see that this disjointness of the
supports of $\mu(\lambda|\psi)$ and $\xi(\psi^{\perp}|\lambda)$ will
also imply (\ref{mu_orthog_disjoint}).
\end{proof}
\\

We can furthermore deduce two simple relations for the supports of
indicator functions associated with a PVM measurement.

\begin{lemma}
The set of indicator functions $\left\{\xi(k|\lambda)\right\}_k$
associated with elements $\left\{|k\rangle\langle{k}|\right\}_k$ of
a PVM measurement must have supports completely spanning the
system's ontic state space $\Lambda$,

\begin{equation}
\bigcup_k\text{Supp}(\xi(k|\lambda))=\Lambda.
\label{ind_fns_span_lambda}
\end{equation}
\label{LEM:xi_pvm_span_lambda}
\end{lemma}

\begin{proof}
This Lemma follows directly from (\ref{ind_fns_sum_one}), which
encodes the quantum mechanical requirement that a PVM should always
exhibit \textit{some} outcome. (\ref{ind_fns_span_lambda}) ensures
that an ontological model will predict some PVM outcome, no matter
what ontic state describes a system.
\end{proof}
\\

We can in fact see that the supports of indicator functions
associated with elements of a PVM not only span $\Lambda$ (as in
(\ref{ind_fns_span_lambda})), but - if the indicator functions are
idempotent - furthermore partition it,

\begin{lemma}
A set of deterministic indicator functions
$\left\{\xi(k|\lambda)\right\}_k$ associated with elements
$\left\{|k\rangle\langle{k}|\right\}_k$ of a PVM measurement must
have mutually disjoint supports,
\begin{equation}
\text{Supp}(\xi(k|\lambda))\cap\text{Supp}(\xi(l|\lambda))=\emptyset.
\label{ind_fns_disjoint_pvm}
\end{equation}
Where $k$ and $l$ label any two elements of the PVM measurement.
\label{LEM:xi_pvm_disjoint}
\end{lemma}

\begin{proof}
This result follows from the quantum mechanical prediction that we
should only ever obtain \textit{one} outcome in any complete PVM
measurement. The disjointness of the indicator functions associated
with different PVM outcomes is necessary to ensure that there is no
$\lambda\in\Lambda$ that would yield more than one outcome for the
PVM.
\end{proof}
\\

There are also several quantum mechanical relations between density
operators and POVM elements that must carry over to relations
between epistemic states and indicator functions, as was noted in
\cite{spekkens_con}. Firstly, consider different convex
decompositions of a density operator.

\begin{lemma}
If a density operator can be prepared according to a convex
decomposition $\rho=\sum_ip_i|\psi_i\rangle\langle\psi_i|$
(corresponding to configuring a measurement device according to some
setting $S_{\mathcal{P}}$) then the epistemic state associated with
$\rho$ when prepared in this way must satisfy a similar relation,
\begin{equation}
\mu(\lambda|\rho,S_{\mathcal{P}})=\sum_ip_i\mu(\lambda|\psi_i).
\end{equation}
\label{LEM:mu_convex_relation}
\end{lemma}

\begin{proof}
This follows from a purely operational argument. We wish
$\mu(\lambda|\rho,S_{\mathcal{P}})$ to give the probability of the
ontic state of $\mathcal{S}$ being $\lambda$ given that
$\mathcal{P}$ was configured with a setting $S_{\mathcal{P}}$. Now
$S_{\mathcal{P}}$ corresponds to a convex decomposition wherein,
with probability $p_i$, the chance of obtaining $\lambda$ is given
by the probability with which we would expect to find $\lambda$ if
the system was described by quantum state $\psi_i$. But the
ontological model's prediction for the latter probability is just
$\mu(\lambda|\psi_i)$ and so the overall probability of obtaining
$\lambda$ is given by $\sum_ip_i\mu(\lambda|\psi_i)$, as stated
above.
\end{proof}
\\

Similarly, we can show that any convex structure of POVM
measurements must carry over to the ontological model description of
measurements,

\begin{lemma}
If a POVM can be prepared by probabilistically performing one of a
set of PVM measurements, so that a particular effect $E$ from the
POVM can be implemented as $E=\sum_ip_i|\psi_i\rangle\langle\psi_i|$
(corresponding to a setting $S_{\mathcal{M}}$ of a measurement
device $\mathcal{M}$) then the indicator function associated with
$E$ when implemented in this way must satisfy,
\begin{equation}
\xi(E|\lambda,S_{\mathcal{M}})=\sum_ip_i\xi(\psi_i|\lambda).
\end{equation}
\label{LEM:indfn_convex_relation}
\end{lemma}

\begin{proof}
The proof of this Lemma follows from an operational argument similar
to the proof given for Lemma \ref{LEM:mu_convex_relation}.
Performing the measurement $E=\sum_ip_i|\psi_i\rangle\langle\psi_i|$
implies an operational procedure wherein we choose a label $i$
according to the probability distribution $\left\{p_i\right\}_i$ and
then implement the associated rank one measurement
$P_i=|\psi_i\rangle\langle\psi_i|$ by performing the PVM
$\left\{P_i,\openone-P_i\right\}$. Now one can ask how we might
write the probability $\xi(E|\lambda,S_{\mathcal{M}})$ for obtaining
the outcome associated with $E$ given that the ontic state of the
system was $\lambda$. Had we performed a rank one projective
measurement $P_i$ then the probability of getting a positive outcome
would be $\xi(\psi_i|\lambda)$. Since $E$ corresponds to
implementing $P_i$ with probability $p_i$ then it follows from
elementary probability theory that
$\xi(E|\lambda,S_{\mathcal{M}})=\sum_i{p}_i\xi(\psi_i|\lambda)$,
which is the desired result.
\end{proof}

\end{appendix}

\end{document}